\documentclass[12pt]{article}

\pdfoutput=1

\usepackage{amsmath,amssymb,amsfonts,amscd,mathrsfs}
\usepackage{xcolor}
\definecolor{darkblue}{rgb}{0.1,0.1,.7}
\usepackage[colorlinks, linkcolor=darkblue, citecolor=darkblue, urlcolor=darkblue, linktocpage]{hyperref} 
\usepackage{geometry}
\usepackage{tablefootnote,colortbl}

\usepackage{braket}

\usepackage{cancel}
\usepackage{arcs}

\usepackage{arydshln}
\usepackage{booktabs,multirow}
\usepackage{amssymb}
\usepackage{fancyvrb}

 \usepackage[square, comma, compress,numbers]{natbib}
\usepackage{subcaption}

\definecolor{myorange}{RGB}{199,146,32}

\definecolor{Gray1}{gray}{0.97}
\definecolor{Gray2}{gray}{0.9}
\definecolor{LightCyan}{rgb}{0.88,1,1}
\definecolor{blu}{rgb}{0,0,1}

\usepackage{ragged2e}
\usepackage{array}
\newcolumntype{L}[1]{>{\raggedright\let\newline\\\arraybackslash\hspace{0pt}}m{#1}}
\newcolumntype{C}[1]{>{\centering\let\newline\\\arraybackslash\hspace{0pt}}m{#1}}
\newcolumntype{R}[1]{>{\raggedleft\let\newline\\\arraybackslash\hspace{0pt}}m{#1}}

\usepackage{dsfont} 
\usepackage{tcolorbox}
\usepackage{booktabs}

\usepackage{titlesec}
\titleformat*{\section}{\large\bfseries}
\titleformat*{\subsection}{\normalsize\bfseries}
\titleformat*{\subsubsection}{\normalsize\it}
\titleformat*{\paragraph}{\normalsize\bfseries}
\titleformat*{\subparagraph}{\normalsize\bfseries}

\newcommand{\reef}[1]{(\ref{#1})}

\newcommand{\beq}{\begin{equation}} 
\newcommand{\eeq}{\end{equation}}
 
\def\nn{\nonumber}

\def\geq{\geqslant}
\def\leq{\leqslant}
\newcommand{\diffop}[2]{\ifthenelse{\equal{#2}{1}}{\frac{\mrm{d}}{\mrm{d} #1}}{\frac{\mrm{d}^#2}{\mrm{d} #1^#2}}}

\newcommand{\mrm}[1]{{\mathrm #1}}

\newcommand{\im}{\text{Im}\, }
\newcommand{\re}{\text{Re}\, }

\usepackage[normalem]{ulem}

\newcommand{\be}{\begin{equation}}
\newcommand{\ee}{\end{equation}}
\def\bea#1\eea{\begin{align}#1\end{align}}

\usepackage{accents}
\newlength{\dhatheight}

\numberwithin{equation}{section}
\usepackage{tocloft}
\begin{document}

\vspace*{-.6in} \thispagestyle{empty}
\begin{flushright}
\end{flushright}
\vspace{1cm} {\Large
\begin{center}
\textbf{
A Geometric View on Crossing-Symmetric \\[.2cm]  Dispersion Relations
}
\end{center}}
\vspace{1cm}
\begin{center}

{\bf  Joan Elias Mir\'o$^{a}$, Andrea Guerrieri$^{b}$, Mehmet As{\i}m G\"{u}m\"{u}\c{s}$^{c}$, Ahmadullah Zahed$^{a}$} \\[1cm] 
 {$^a$ The Abdus Salam ICTP,    Strada Costiera 11, 34135, Trieste, Italy  \\ 
 $^b$Department of Mathematics, City St. George’s, Univ. of London,
Northampton Square, EC1V 0HB, London, UK\\
 $^c$LAPTh, CNRS et Univ. Savoie Mont-Blanc, 9 Chemin de Bellevue, F-74941 Annecy, France
}
\vspace{1cm}

\abstract{ 

We introduce a general framework for constructing dispersion relations using crossing-symmetric variables, leading to infinitely many distinct representations of the $2\!\to\!2$ scattering amplitude of identical scalars. Classical formulations such as the Auberson–Khuri crossing-symmetric dispersion relations (CSDRs), the Mahoux–Roy–Wanders relations, and the local CSDR, as well as fixed-$t$ dispersion relations emerge as special cases. Within this setting we re-derive the \emph{null constraints} from a geometric perspective. Finally, we present, for the first time, an explicit extension of Roy-like equations that remain valid at arbitrarily high energies, relying only on the rigorously established analyticity domain of scattering amplitudes.

}

\vspace{3cm}
\end{center}

 \vfill
 {
  \flushright
 \today 
}

\newpage 

\setcounter{tocdepth}{1}

{
\tableofcontents
}
 
 

\section{Introduction}

Exploring the analytic structure of scattering amplitudes often reveals remarkably rich mathematical properties. One such framework is that of Crossing-Symmetric Dispersion Relations (CSDRs). Originally introduced in the 1970s~\cite{Auberson:1972prg,Mahoux:1974ej} and further developed in~\cite{Atkinson:1974ev}, the crossing-symmetric formalism has recently found renewed applications in both scattering amplitudes and conformal field theory~\cite{Sinha:2020win,Gopakumar:2021dvg}. 
Recent developments span a wide range of contexts: meromorphic amplitudes~\cite{Saha:2024qpt,Bhat:2024agd,Bhat:2025zex}, gravitational theories~\cite{deRham:2022gfe,Chang:2025cxc,Pasiecznik:2025eqc}, conformal field theory~\cite{Alday:2022xwz,Bhat:2023ekh,Bissi:2022fmj}, spinning scattering processes~\cite{Chowdhury:2021ynh}, as well as pion scattering and chiral perturbation theory~\cite{Zahed:2021fkp,Li:2023qzs} and its connections with geometric function theory~\cite{Haldar:2021rri,Raman:2021pkf}.

In this work, we revisit CSDRs from the perspective of the non-perturbative $3+1$-dimensional $S$-matrix bootstrap program~\cite{Paulos:2017fhb,Guerrieri:2018uew,Guerrieri:2020bto,Hebbar:2020ukp,Guerrieri:2021ivu,He:2021eqn,Chen:2022nym,EliasMiro:2022xaa,Guerrieri:2022sod,Tourkine:2023xtu,Acanfora:2023axz,Gumus:2024lmj,Bhat:2023puy,Guerrieri:2024jkn,deRham:2025vaq,Correia:2025uvc}. 
Our investigation is motivated by a longstanding obstacle in the dual formulation of the fixed-$t$ bootstrap. Originally developed in~\cite{Lopez:1974cq,Lopez:1976zs}, this approach has recently been revived for gaped scalar theories~\cite{Guerrieri:2021tak}, extended to theories with internal symmetries~\cite{EliasMiro:2023fqi}, and applied to glueball scattering in pure Yang--Mills~\cite{Guerrieri:2023qbg}. A key limitation in all these cases is the restricted range of energies where non-perturbative unitary can be imposed: typically $4m^2 \leq s \leq 12m^2$.~\footnote{More generally, $\Lambda^2 \leq s \leq 4\Lambda^2 - 4m^2$, where $\Lambda$ denotes the first threshold singularity. In weakly coupled EFTs, $\Lambda$ can often be identified with the UV cutoff~\cite{EliasMiro:2023fqi}, whereas for strongly coupled theories one typically has $\Lambda=2m$.} This is not a numerical artifact but a fundamental constraint imposed by the feasibility of the dual optimization problem. 

\begin{figure}[h!] 
    \centering 
    \includegraphics[width=.9\textwidth]{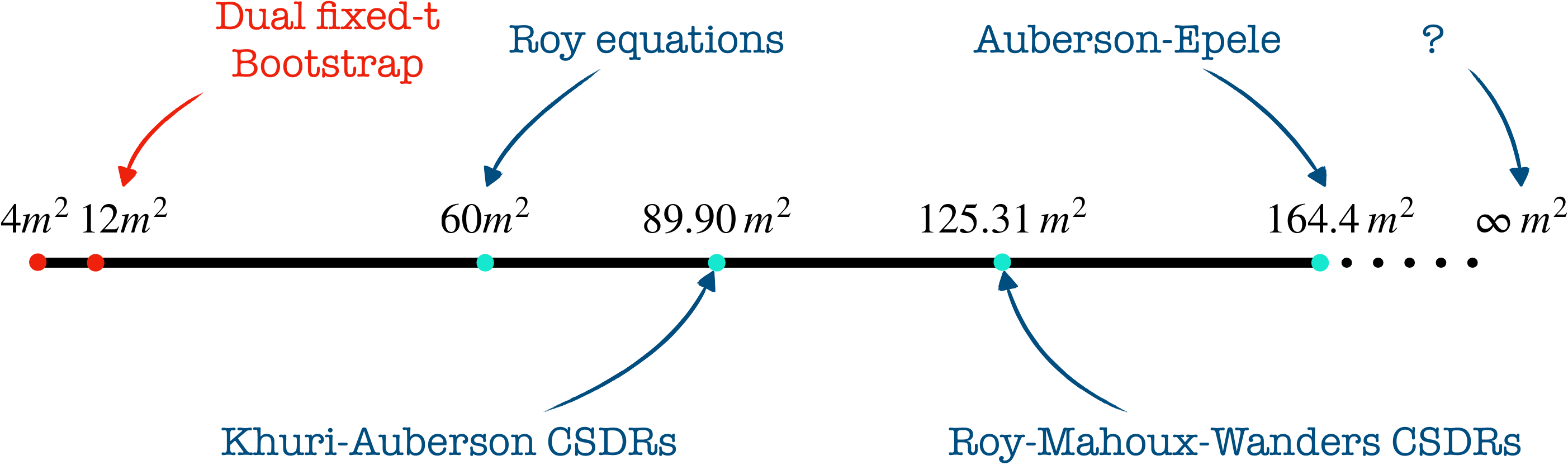} 
	\caption{Comparison of the validity domains for various dispersive approaches. The values indicate the maximum $s$ compatible with the proven analytical properties of amplitudes. The last known extension in the literature is due to Auberson and Epele~\cite{Auberson:1974in}.}
\label{plot_cutoffs}
\end{figure}

As shown in Fig.~\ref{plot_cutoffs}, the validity domain of current dual bootstrap implementations falls short of what is allowed by rigorous fixed-$t$ dispersion relations. This mismatch is especially unsatisfactory in realistic applications such as $\pi\pi$ scattering, where the resonance spectrum extends far beyond $12m_\pi^2$. Moreover, even if the bootstrap could be pushed further, one would eventually encounter the intrinsic limitation of the partial wave expansion. For fixed-$t$ dispersion relations, this corresponds to $s=60m^2$, the conventional limit of Roy equations~\cite{Roy:1971tc}.

Remarkably, this barrier was already surpassed in the 1970s by two independent approaches. Auberson and Khuri (AK)~\cite{Auberson:1972prg}, using crossing-symmetric variables, extended the validity to $s=89.90m^2$. Soon after, Mahoux, Roy, and Wanders (MRW)~\cite{Mahoux:1974ej} refined the AK construction and reached $s=125.31m^2$. They also emphasized that this extension cannot be achieved with a single dispersion relation: different formulas are required in overlapping intervals, each involving independent subtraction constants. The last significant enlargement was achieved by Auberson and Epele~\cite{Auberson:1974in}, who reached $s=164.4m^2$. To our knowledge, however, no simple or universal set of dispersion relations exists that covers the entire physical region.~\footnote{See also Auberson and Ciulli~\cite{Auberson:1977re} for a related attempt.}

The main goal of this paper is to revisit CSDRs from a general geometric perspective and to address the longstanding problem of constructing dispersion relations valid at all energies. We clarify the origin of the validity ranges of existing relations (which appear puzzling in Fig.~\ref{plot_cutoffs}) and introduce the geometric language of foliations into this context. Building on this framework, we construct a new family of CSDRs that extend Roy-like equations to cover the full range of physical energies.

The paper is organized as follows. In Section~\ref{kinematics_xy} we introduce the standard crossing-symmetric variables \( x(s_1,s_2) \) and \( y(s_1,s_2) \) to solve for crossing symmetry. In Sections~\ref{lincase} and~\ref{sec:nonlinear_leaves}, we write dispersion relations in \( x \), on a one-parameter family of foliations \( y = f_\alpha(x) \), involving both linear and non-linear functions of \( x \). Unitary is not manifest in this representation. We construct the CSDRs in Section~\ref{pullback_linear} by pulling back the relation \( y = f_\alpha(x) \) to the physical Mandelstam variables \( s, t, u \), where unitary is manifest. The resulting expressions are~\eqref{eq:final_csdr} and~\eqref{eq:pulled_disp_on_non_linear_leaf}. In Section~\ref{nullc} we analyze the singularity structure of the foliations, leading to null constraints. Section~\ref{roystuff} reviews the analytic properties of the amplitude and the crossing-symmetric representation, and concludes with the implications of analyticity for Roy-like equations defined on \( y = f_\alpha(x) \). We extend the domain of validity using a `homogeneous foliation", yielding~\eqref{royex} 
for all \( s \in [4,\infty) \). Finally, we conclude with an outlook on possible continuations of this research.

\section{Kinematical space using crossing-symmetric variables}
\label{kinematics_xy}

We consider the two-to-two scattering amplitude \( M(s, t) \) of identical scalar particles of mass \( m \), without cubic self-interactions. The amplitude is totally symmetric in the Mandelstam variables, which satisfy the constraint \( s + t + u = 4m^2 \). 
To simplify the algebra, we introduce the shifted variables 
\begin{equation}
s_1 = s - \tfrac{4m^2}{3}, \quad s_2 = t - \tfrac{4m^2}{3}, \quad s_3 = u - \tfrac{4m^2}{3},
\end{equation}
so that the symmetric point \( s = t = u = 4m^2/3 \) is mapped to the origin, and momentum conservation reads \( s_1 + s_2 + s_3 = 0 \). 

\begin{figure}[t!] 
\centering 
\includegraphics[width=1\textwidth]{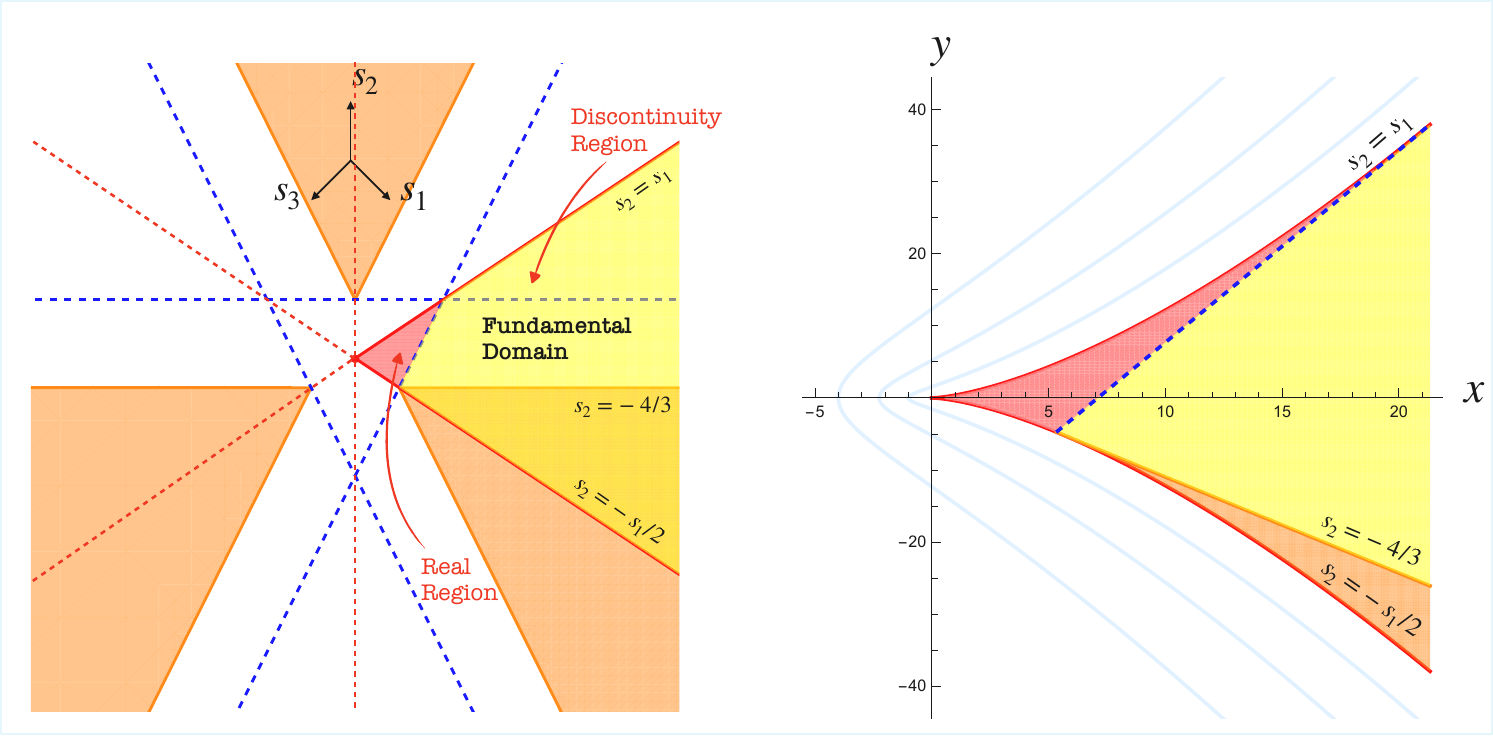}
\caption{
\textbf{Left:} Fundamental domain projected onto real Mandelstam variables, delimited by the red lines and defined by the inequalities \( -s_1/2 \leq s_2 \leq s_1 \), with \( s_1 \geq 0 \). Dashed blue lines correspond to \( s_i = 8/3 \). The amplitude develops a discontinuity in \( s_1 \) over the orange/yellow region. The orange region is the physical scattering domain \( s_1 \geq 8/3 \), \( 0 \leq \cos\theta \leq 1 \). \textbf{Right:} Image of the fundamental domain in real crossing-symmetric variables \( (x, y) \). The region complementary to the fundamental domain where \( x, y \) are real can be parametrized by the lines \( s_2 = -s_1/2 + 2i \tau \), with \( \tau \in \mathbb{R} \), or equivalently 
\( y = \tfrac{2}{3\sqrt{3}} \sqrt{x + \tau^2} \, (x + 4\tau^2) \), shown in light blue.
}
\label{fig_geometry}
\end{figure}

In Figure~\ref{fig_geometry} (left), we visualize the kinematical space on the \emph{Mandelstam plane} defined by the variables \( \{s_1, s_2, s_3\} \) under the constraint of momentum conservation.  In what follows, we define the amplitude \( M(s_1, s_2) \) on this shifted plane and set the mass scale to unity, \( m = 1 \).

Crossing symmetry identifies the values of the amplitude at 
$(s_1,s_2,s_3)$ with those at any permutation of the three variables. 
As a result, the Mandelstam plane is partitioned into six equivalent 
wedges related by permutations of the $s_i$. For definiteness, we work in 
the wedge defined by the ordering $u \leq t \leq s$, which is equivalent to 
\begin{equation}
-\tfrac{s_1}{2} \leq s_2 \leq s_1 .
\label{fund_domain}
\end{equation}
In Figure~\ref{fig_geometry} (left), the dashed and solid red lines indicate 
the boundaries between wedges, with the chosen \emph{fundamental domain}
highlighted in solid red.
Within this region several subdomains are of particular importance.
First, a small red triangle with $s_i \leq 8/3$ lies entirely below 
the two-particle threshold in all channels, so the amplitude there is 
real and analytic.  
Second, the domain also contains half of the physical $s_1$--channel 
region (shaded orange), corresponding to $s_1 \geq 8/3$ and 
$s_2 \in [-s_1/2,-4/3]$, or equivalently to scattering angles 
$\cos\theta \in [0,1]$.  
Beyond this range the amplitude develops a branch cut in $s_1$ even for 
unphysical values of $\cos\theta$; this region is shown in yellow.  
The yellow domain also supports a non-vanishing double discontinuity, 
which we omit in the discussion below as it will not play a role in 
this section.

To eliminate the redundancy among wedges, we introduce the \emph{crossing-symmetric variables}~\cite{Wanders:1996}
\begin{equation}
x = -(s_1 s_2 + s_2 s_3 + s_3 s_1), \qquad y = - s_1 s_2 s_3,
\label{eq:cross_symm_vars}
\end{equation}
and define the amplitude in these variables as \( \mathcal{M}(x, y) \equiv M(s_1(x,y), s_2(x,y)) \). This mapping identifies all six wedges of the Mandelstam plane.

When $s_1,s_2\in\mathbb{R}$, the fundamental domain (and its images) is mapped to the
region bound by
\begin{equation}
27y^2 \leq 4x^{3}, \qquad x \geq 0.
\label{physical_in_xy}
\end{equation}
In Figure~\ref{fig_geometry} (right), the boundary of this region is shown in red.
This inequality follows by rewriting \eqref{fund_domain} in the new variables. 
The symmetric point again maps to the origin, \( (x,y) = (0,0) \). The various colored regions here match their counterpart on the figure on the left.
The complementary region of \eqref{physical_in_xy} in the real $(x,y)$-plane corresponds to kinematical configurations where at least one Mandelstam variable is complex.

\vspace{1em}

The variables 
$(x,y)$ can also be understood algebraically. Consider the cubic polynomial
\begin{equation}
P(\sigma) \equiv (\sigma- \sigma_1)(\sigma - \sigma_2)(\sigma - \sigma_3) = \sigma^3 - \Sigma \, \sigma^2 - x \, \sigma + y
\end{equation}
with \( \Sigma = \sigma_1 + \sigma_2 + \sigma_3 = 0 \) by momentum conservation. 
The polynomial \( P(\sigma) \), by construction, is manifestly invariant under permutations of its roots. 
Therefore, these three roots parametrize the orbifold \( \mathcal{H}_\Sigma / S_3 \), where \( \mathcal{H}_\Sigma \subset \mathbb{C}^3 \) is the two-dimensional complex sub-manifold of triples satisfying \( \sigma_1 + \sigma_2 + \sigma_3 = 0 \), and $S_3$ is the symmetric group of three elements.   
The map between the crossing-symmetric variables \( (x, y) \) and the unordered roots \( \{ \sigma_i \} \in \mathcal{H}_\Sigma / S_3 \) is invertible, unlike the map between $\{s_i\}$ and $x,y$. 
The variables $x$ and $y$ play then the role of holomorphic coordinates for the orbifold.  

To relate $x,y$ to the original Mandelstam variables, one must solve the system of equations
\[
y = \sigma_i(x-\sigma_i^2), \quad i = 1,2,3,
\]
which indeed yields six different solutions, as expected from full crossing symmetry. 

The discriminant~\footnote{The discriminant $\Delta$ of an $n$-th degree polynomial is the product of the squares of the differences of the polynomial roots $r_i$, namely $\Delta = \prod_{i<j}^{n} (r_i - r_j)^2$, also known as the square of the Vandermonde polynomial.}
of the polynomial \( P \) is given by $\Delta(x,y) = 4x^3 - 27y^2$.
This quantity vanishes when two or more roots coincide on the boundary \reef{physical_in_xy}. When \( \Delta \neq 0 \), the roots are distinct.
If \( \Delta > 0 \), all three roots are real, which corresponds to the region \reef{physical_in_xy} as expected. If \( \Delta < 0 \), there is one real root and a pair of complex conjugates.\footnote{Note that the roots can be obtained through Cardano's trigonometric formula $\sigma_k=2\sqrt{\frac{x}{3}} \cos\left(\frac{\theta}{3}+ \frac{2\pi k}{3}\right)$ where $(\cos \theta)^2={(27y^2)/{(4x^3)}}$.}

Since, in general, at least one root is real, we label it \( \sigma \), and express the other two as
\begin{equation}
\tau^\pm(\sigma) = -\frac{\sigma}{2} \pm \frac{1}{2} \sqrt{4x - 3\sigma^2} = -\frac{\sigma}{2} \pm \sqrt{\sigma^2 + \frac{4y}{\sigma}},
\label{other_roots}
\end{equation}
When \( (x,y) \) lies within the region~\eqref{physical_in_xy}, the square root is real and all roots are real. Otherwise, \( \tau^\pm \) are complex conjugates. 
We can associate any of these roots to a Mandelstam invariant, and this choice determines the wedge of the physical kinematical space we use to represent the amplitude.
For instance, we will assign $s_1=\sigma$, $s_2=\tau^+$, and $s_3=\tau^-$ so that the region $\Delta>0$ is mapped to the fundamental domain depicted in Figure \ref{fig_geometry}.
 
\section{Dispersion relations on linear leaves}
\label{lincase}

 The goal of this section is to show that, when expressed in crossing-symmetric variables, it becomes straightforward to apply Cauchy's theorem and derive infinite families of dispersive representations for the amplitude \( \mathcal{M}(x,y) \) as a function of the complex variable \( x \). To reduce the problem to a single complex variable, we fix a relation between \( x \) and \( y \). For simplicity, we consider a linear relation of the form
 \begin{equation}
 y = a(x - x_0),
\label{foliation_linear}
\end{equation}
where \( a \) and \( x_0 \) are complex parameters.

\begin{figure}
    \centering
    \includegraphics[width=\linewidth]{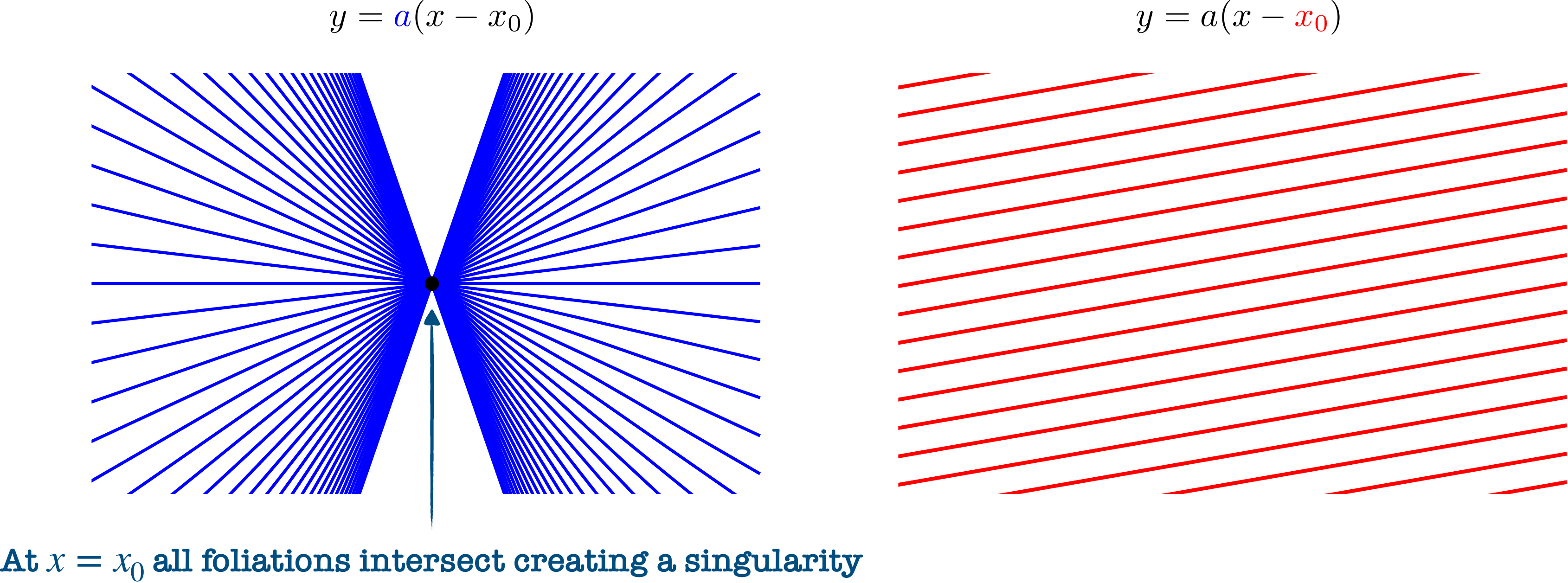}
    \caption{\textbf{Left:} Foliation of the \( (x,y) \) plane generated by varying the slope \( a \) at fixed \( x_0 \). All leaves intersect at the point \( (x_0, 0) \). \textbf{Right:} Parallel foliation obtained by fixing \( a \) and varying \( x_0 \).}
    \label{fig:linear_foliations}
\end{figure}

The relation~\eqref{foliation_linear} defines a complex straight line, i.e. a one-dimensional sub-manifold of the two-dimensional complex space \( \mathbb{C}^2_{(x,y)} \). 
By fixing \( x_0 \) and varying the slope \( a \), we obtain a one-parameter family of such lines that cover the space—see Fig.~\ref{fig:linear_foliations} (left). 
Each line is called a \emph{leaf} of the foliation. 
A point is \emph{regular} if it belongs to a unique leaf; for example, any point with \( x \neq x_0 \). 
In contrast, all leaves intersect at the common point \( (x_0,0) \), which is therefore a \emph{singular point} of the foliation. 
We will return to the classification of such singularities in Section~\ref{nullc}.

Alternatively, we can fix \( a \) and vary \( x_0 \), producing a foliation of parallel leaves, which appears free of singularities. However, as we also discuss in Section~\ref{nullc}, these leaves accumulate at infinity, and thus there is a singularity at the point at infinity. 

Once a leaf is chosen—i.e., once \( a \) and \( x_0 \) are fixed—we can regard \( y \) as a function of \( x \), and define the amplitude restricted to the leaf as
\[
\mathcal{M}_{a,x_0}(x) \equiv \mathcal{M}(x, a(x - x_0)).
\]
If \( \mathcal{M}_{a,x_0}(x) \) is analytic in a neighborhood of \( x \), we can apply Cauchy's theorem:
\begin{equation}
\mathcal{M}_{a,x_0}(x) = \frac{1}{2\pi i} \oint_x dx' \, \frac{\mathcal{P}(x') \mathcal{M}_{a,x_0}(x')}{x' - x},
\label{cauchy_foliation}
\end{equation}
where \( \mathcal{P}(x') \) is a regulator function that ensures convergence by suppressing the arc at infinity; a suitable choice is
$
\mathcal{P}(x') = \frac{x - x_s}{x' - x_s}
$,
with \( x_s \) an arbitrary subtraction point.\footnote{Indeed, the Froissart bound states \( \mathcal{M}_{a,x_0}(x(s_1, s_2 \leq 0)) \leq \pi s_1 \log^2(s_1/\text{const}) \) as \( s_1 \to \infty \), and since \( x \sim s_1^2 \), the amplitude grows no faster than \( \sqrt{x} \log^2 x \).} 
Assuming that \( \mathcal{M}_a(x') \) has no singularities in the complex \( x' \)-plane aside from the physical \( s_1 \)-channel cut, we can deform the contour and isolate the discontinuity:
\begin{equation}
\mathcal{M}_{a,x_0}(x) = \mathcal{M}_{a,x_0}(x_s) + \frac{1}{\pi} \int_{x_\text{th}}^\infty dx' \, \frac{x - x_s}{x' - x_s} \, \frac{\mathcal{A}_{a,x_0}(x')}{x' - x},
\label{eq:disp_on_linear_leaf}
\end{equation}
where the discontinuity across the cut is defined as
\begin{equation}
\mathcal{A}_{a,x_0}(x) \equiv \text{disc}_x \mathcal{M}_{a,x_0}(x) = \lim_{\epsilon \to 0} \frac{1}{2i} \left[ \mathcal{M}_{a,x_0}(x + i\epsilon) - \mathcal{M}_{a,x_0}(x - i\epsilon) \right].
\label{asin}
\end{equation}

The lower limit of integration, \( x_\text{th} \), corresponds to the point where the leaf intersects the threshold line 
\( y = \frac{8}{3} \left(x - \tfrac{64}{9} \right) \), i.e., the boundary of the physical region. Solving
$
a(x - x_0) = \frac{8}{3} \left(x - \tfrac{64}{9} \right)
$,
yields
\begin{equation}
x_\text{th} = \frac{a x_0 - (8/3)^3}{a - 8/3}.
\end{equation}

\begin{figure}[t!] 
\centering 
\includegraphics[width=1\textwidth]{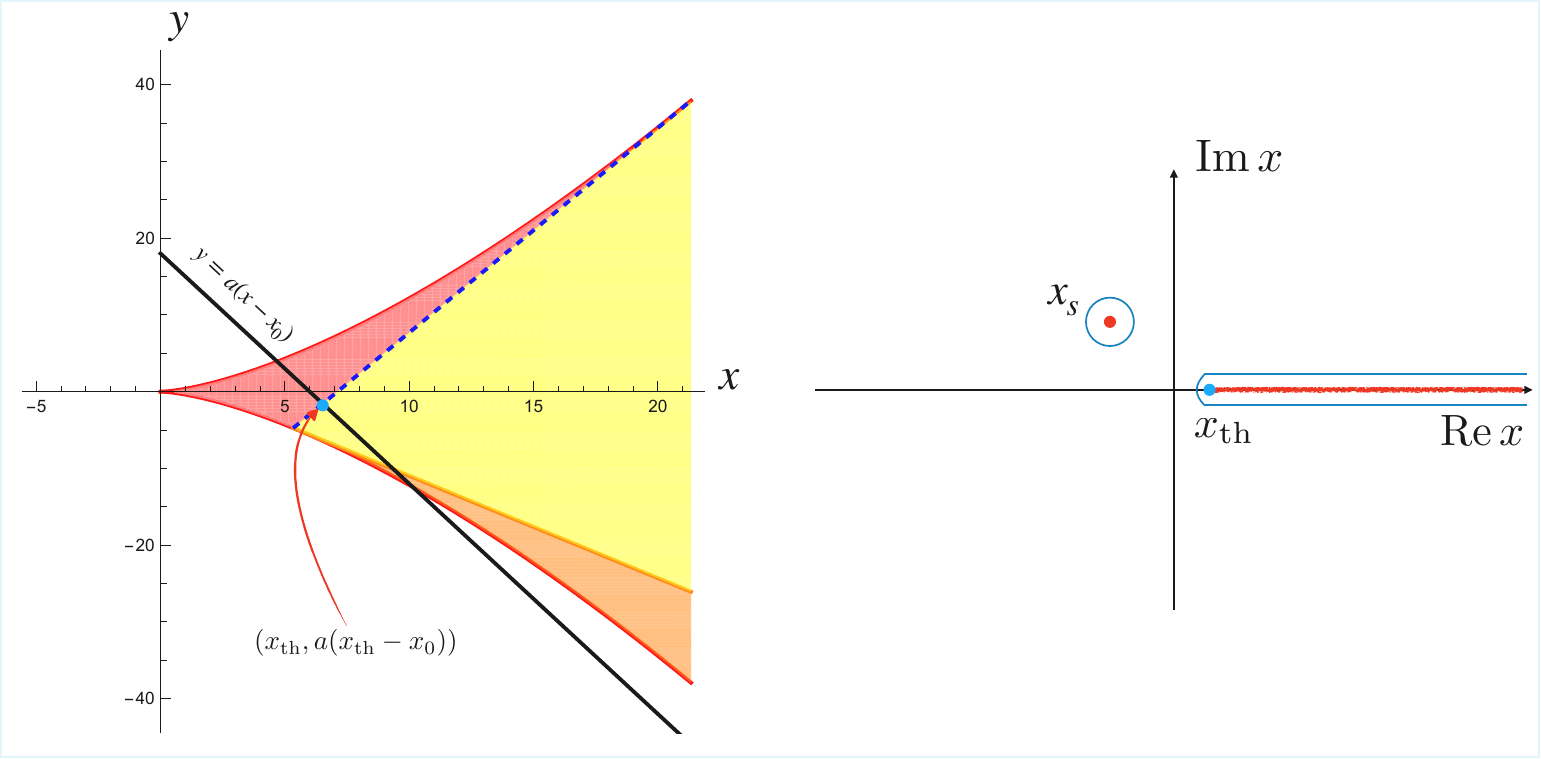} 
\caption{
\textbf{Left:} A linear foliation \( f(x) = a(x - x_0) \) in the \( (x, y) \) plane. When the leaf intersects the threshold line 
$y = \tfrac{8}{3}\left( x-\tfrac{64}{9} \right)$
(dashed blue), the amplitude develops a branch cut starting at \( x_\text{th} = \tfrac{a x_0 - (8/3)^3}{a - 8/3} \). \textbf{Right:} Complex \( x \)-plane showing the deformed Cauchy contour. The relevant singularities are the branch point at \( x_\text{th} \) and the subtraction point \( x_s \).}
\label{fig_disp_in_x}
\end{figure}

Figure~\ref{fig_disp_in_x} illustrates this construction. In the \textbf{left panel}, we show the foliation \( y = a(x - x_0) \) in the \( (x,y) \)-plane, and its intersection with the physical threshold line (dashed blue). In the \textbf{right panel}, we depict the integration contour in the complex \( x \)-plane used in the derivation of Eq.~\eqref{eq:disp_on_linear_leaf}. The only relevant singularities are the two-particle threshold \( x_\text{th} \) and the subtraction point \( x_s \); the regulator ensures that we can drop the arc at infinity.

Finally, the reader might ask whether the dispersive representation derived here is rigorously justified. So far, we have assumed \emph{maximal analyticity}, but in Section~\ref{roystuff} we will revisit the rigorous domain of analyticity derived by Martin~\cite{Martin:1965jj,Martin:1966zsy}, which imposes additional constraints on the admissible foliations—and hence on the allowed values of \( a \) and \( x_0 \).

\subsection{Pulling-back to Mandelstam variables}
\label{pullback_linear}

Equation~\eqref{eq:disp_on_linear_leaf} defines a CSDR by construction. To relate it to the amplitude for physical energies and scattering angles and to analyze how unitary manifests itself, we must express it in terms of the standard Mandelstam invariants. More precisely, we use the variables \( \sigma_i \) introduced earlier to parametrize the kinematical orbifold \( \mathcal{H}_\Sigma/S_3 \).

Recall that \( (x, y) \) and \( \{\sigma_i\} \) are related via a symmetric polynomial equation \( P(\sigma) = 0 \), which implicitly defines a one-to-one map between the crossing-symmetric variables and the unordered Mandelstam invariants. For the foliation \( y = a(x - x_0) \), this equation becomes:
\begin{equation}
P(\sigma) = \sigma^3 - x \sigma + a(x - x_0) = 0,
\end{equation}
whose solution \( x \) we can be expressed as a rational function of \( \sigma \):
\begin{equation}
x(\sigma) = \frac{a x_0-\sigma^3}{a-\sigma}.
\label{eq:change_in_x}
\end{equation}

We can now rewrite the dispersive representation~\eqref{eq:disp_on_linear_leaf} in terms of \( \sigma \), via a simple change of variables $x^\prime(\sigma)=\frac{a x_0-\sigma^3}{a-\sigma}$,
\begin{equation}
\mathcal{M}_{a,x_0}(x) = \mathcal{M}_{a,x_0}(x_s) + \frac{1}{\pi} \int_{\frac{8}{3}}^\infty \frac{dx'(\sigma)}{d\sigma} \, \frac{x - x_s}{x'(\sigma) - x_s} \, \frac{\mathcal{A}_{a,x_0}(x'(\sigma))}{x'(\sigma) - x} \, d\sigma.
\label{eq:disp_on_sigma}
\end{equation}

Since crossing has been implemented through symmetrization, the variable \( \sigma \) can be interpreted as any of the three Mandelstam variables \( s_1, s_2, s_3 \). To make contact with the physical region and the partial wave expansion within, we choose \( \sigma = s_1 \), so that the discontinuty of the amplitude is evaluated in the \( s_1 \)-channel. 
Next, to express the variables \( x \) and \( x_s \) in terms of \( (s_1, s_2) \), we equate the symmetric definition of \( x \) to the expression obtained from the foliation:
\begin{equation}
x(s_1, s_2) = s_1^2 + s_1 s_2 + s_2^2 = \frac{a x_0-s_1^3}{a-s_1}.
\label{eq:x_match}
\end{equation}
Solving this relation yields two solutions for \( s_2 \):
\begin{equation}
s_2^\pm(s_1) = -\frac{s_1}{2} \pm \frac{1}{2} \sqrt{ \frac{s_1^3 + a(3s_1^2-4x_0)}{s_1-a} }.
\label{eq:s2_plusminus}
\end{equation}

These equations can now be used to evaluate the amplitude and its discontinuity in terms of the original Mandelstam variables. In particular, setting \( s_2 = s_2^+(s_1) \) (setting $s_2=s_2^-(s_1)$ yields identical formulas), we compute:
\begin{itemize}
\item the value of \( x(s_1, s_2^+(s_1)) \equiv x(s_1,s_2)\),
\item the value of the subtraction point \( x_s(s_1^0, s_2^+(s_1^0)) \equiv x_s(s_1^0,s_2^0)\),
\item the derivative \( dx'/d\sigma \).
\end{itemize}
For the ease of notation, we   define
\begin{itemize}
    \item $ M(s_1,s_2)\equiv \mathcal{M}_{a,x_0}(x(s_1,s_2))$
    \item  $ A(\sigma, s_2^+(\sigma))\equiv \mathcal{A}_{a,x_0}(x^\prime(\sigma))$, is the s-channel discontinuity. This definition uses equations \eqref{eq:x_match} and  \eqref{eq:s2_plusminus}: $x^\prime(\sigma)=x(\sigma,s_2^+(\sigma))$.
\end{itemize}

Then, changing variables in Eq.~\eqref{eq:disp_on_sigma}  leads to the final form of the CSDR in the physical variables: 
\begin{equation}
\boxed{
\begin{aligned}
M(s_1, s_2) - M(s_1^0, s_2^0) &= \frac{1}{\pi} \int_{8/3}^\infty d\sigma \, A(\sigma, s_2^+(\sigma)) \\
&\quad \times \left( \frac{1}{\sigma - s_1} + \frac{1}{\sigma - s_2} + \frac{1}{\sigma - s_3} - \frac{1}{\sigma - s_1^0} - \frac{1}{\sigma - s_2^0} - \frac{1}{\sigma - s_3^0} \right),
\end{aligned}
}
\label{eq:final_csdr}
\end{equation}
where \( s_3 = -s_1 - s_2 \), and \( (s_1^0, s_2^0) \) is the subtraction point lying on the same leaf. The absorptive part of the amplitude can be easily related to partial waves expanding with the appropriate angle
\be
A(\sigma, s_2^+(\sigma))=16\pi\sum_{\ell=0}^\infty(2\ell+1)\im f_\ell(\sigma) P_\ell\left(1+\frac{2 (s_2^+(\sigma)+4/3)}{\sigma-8/3}\right).
\ee

We note  that while \eqref{eq:final_csdr} is explicitly symmetric under permutations of $\{ s_1, s_2, s_3 \}$, it is not valid for arbitrary choices of the Mandelstam variables. Indeed, given a value of $s_1$, the value of $s_2$ is fixed by  $s_2=s_2^+(s_1)$. 
To cover the full kinematical space one needs to vary $a$, thus \eqref{eq:final_csdr} is effectively a one-parameter family of dispersion relations, whose union does cover all physical values of $\{s_1,s_2\}$. As we will see in Section~\ref{nullc},  the requirement that these dispersion relations are mutually compatible, for different values of $a$, leads to regularity constraints, which are often called null constraints in the literature. 

Various dispersions relations known in the literature can be obtained from \eqref{eq:final_csdr} by appropriate choices of the parameters  $\{a,~x_0\}$ and subtraction point $x_s$. Here are a couple of examples:

\paragraph{Mahoux-Roy-Wanders (MRW).} Dispersion relation worked out by Mahoux-Roy-Wanders~\cite{Mahoux:1974ej} considered $f(x)=a(x-x_0)$ for real constants $a$ and $x_0$. The reader can readily compare our Eq.~\reef{eq:final_csdr} (or the alternative form Eq.~\reef{eq:pulled_disp_on_non_linear_leaf} we derive later) to the Eq.~(3.7) of~\cite{Mahoux:1974ej} for a generic subtraction point $x_s$. Note that the subtraction point is in general different from the intercept point $x_0$. When both points coincide, we obtain the dispersion relation~\reef{eq:final_csdr} where $(s_1^{0},s_2^{0},s_3^{0})=(0,2\sqrt{x_0},-2\sqrt{x_0})$, which is then analogous to their Eq.~(3.10).~\footnote{We also found out a typo in \cite{Mahoux:1974ej}, namely, the $x_0$ in their Eq. (3.9) should be replaced with $16x_0$.}

\paragraph{Auberson-Khuri (AK).} The dispersion relations derived by Auberson-Khuri \cite{Auberson:1972prg,Sinha:2020win} are a special case of Eq.~\reef{eq:disp_on_sigma} for $x_0=0$.
For instance, choosing to subtract at $(s_1^{0},s_2^{0},s_3^{0})=(0,0,0)$ yields immediately the Eq.~(2) of \cite{Sinha:2020win}. Note that Auberson-Khuri's paper \cite{Auberson:1972prg} introduces an additional parametric dependence in the Mandelstam variables, given by the functions $s_i(z,a)$, and the dispersion relations are written in the complex $z$-plane for fixed $a$. The formula \reef{eq:final_csdr} recovers the same dispersion relations upon choosing $x_s=x_0=0$, while bypassing the complicated intermediate steps in the $z$-plane.
    
\paragraph{Fixed-$t$ dispersion relations.} The formula \reef{eq:final_csdr} reduces to a fixed-$t$ dispersion relation upon appropriate choices of the parameters. A constant momentum transfer $s_2=\bar{\tau}$ corresponds to the MRW case for $a=\bar{\tau}$ and $x_0=\bar{\tau}^2$. 
The generic subtraction point $x_s$ is then located at $s_2^{0}=\bar{\tau}$ and $s^0_{1,3}=-\bar{\tau}/2\pm\sqrt{x_s{-}3\bar{\tau}^2/4}$.
The subtraction choice $x_s=x_0$, for instance, yields the double subtracted fixed-$t$ dispersion relations in Eq.~(4) of \cite{Jin:1964zza} for equal masses.

\subsection{A new representation with a subtraction constant}

In the previous examples, the subtraction point $x_s$ was generically a function of the parameters $\{a,x_0\}$, i.e. its position depends on the specific leave of the foliation. In the following, we recast the CSDR \eqref{eq:final_csdr} in a different form, where $x_s$ is traded for a single constant $M(0,0)$ across all leaves. Such a form becomes useful when writing Roy-like equations, since the constant term projects only on spin-zero component of the amplitude. In order to achieve this we proceed via the following two steps:

First, let us consider the dispersion relation on the leaf $y=0$, by taking the limit $a \to 0$ in \eqref{eq:final_csdr}. 
Note that points on this leaf map back to the Mandelstam variables $(s_1,s_2,s_3)=(\sqrt{x},0,-\sqrt{x})$ in the fundamental domain where $s_1 \geq s_2 \geq s_3$, and the origin of the $\{x,y\}$ plane is also included in particular. We can choose the origin as the subtraction point, and \eqref{eq:final_csdr} becomes
\begin{equation}
M(\sqrt{x},0) = M(0, 0) + \frac{1}{\pi} \int_{8/3}^\infty d\sigma \, A(\sigma, 0)
\left( \frac{1}{\sigma + \sqrt{x}} + \frac{1}{\sigma - \sqrt{x}} - \frac{2}{\sigma} \right) \, .
\label{eq:y0leaf}
\end{equation}
Next, we consider a dispersion relation on a second leaf $y=a(x-x_0)$ for some $\{a,x_0\}$. We choose the intercept point $(x,y)=(x_0,0)$ as the subtraction point in \eqref{eq:final_csdr}, yielding
\begin{equation}
\begin{aligned}
M(s_1,s_2) &= M(\sqrt{x_0}, 0) + \frac{1}{\pi} \int_{8/3}^\infty d\sigma \, A(\sigma, s_2^+(\sigma)) \\
&\times \left( \frac{1}{\sigma - s_1} + \frac{1}{\sigma - s_2} + \frac{1}{\sigma + s_1 + s_2} - \frac{1}{\sigma + \sqrt{x_0}} - \frac{1}{\sigma - \sqrt{x_0}} - \frac{1}{\sigma} \right) \, .
\label{eq:yax0leaf}
\end{aligned}
\end{equation}
Observe that the intercept of the $x$-axis is a common kinematic point to both leaves, which allows us to replace $M(\sqrt{x_0},0)$ that appears in \eqref{eq:yax0leaf} with the expression given in \eqref{eq:y0leaf} after setting $x=x_0$. Then, we get
\begin{equation}
\boxed{
\begin{aligned}
M(s_1,s_2) &= M(0, 0) + \int_{8/3}^\infty \frac{d\sigma}{\pi} \, \frac{1}{\sigma(\sigma^2 - x_0)} \left[ \, 2x_0 \,  A(\sigma, 0) + (3\sigma^2 - x_0) \, A(\sigma, s_2^+(\sigma)) \, \right] \\
&+ \int_{8/3}^\infty \frac{d\sigma}{\pi} \, A(\sigma, s_2^+(\sigma)) \left( \frac{1}{\sigma - s_1} + \frac{1}{\sigma - s_2} + \frac{1}{\sigma + s_1 + s_2}\right) \, .
\label{eq:roytrick_csdr}
\end{aligned}
}
\end{equation}
In this form, we fix the arbitrary subtraction point $x_s$ in the MRW dispersion relations to be the $(s_1^0,s_2^0,s_3^0)=(0,0,0)$ for any choice of $\{a,x_0\}$. To our knowledge, this is a new representation of the dispersion relations on the linear foliations, where the amplitude is fully determined by the constant $M(0,0)$ and the $s$-channel discontinuity $A(\sigma,s_2^+(\sigma))$. 

\paragraph{Local CSDRs as a limiting case.}   
Next, we show that local CSDR (LCSDR) introduced in \cite{song} is an instance of the linear foliation with   $y=c$ slices for constant $c$. This corresponds to the limit $a \to 0$  with  $-a x_0 = c$  kept fixed in the MRW foliation. First, we note that in this limit \eqref{eq:s2_plusminus} becomes
\begin{equation}
s_2^\pm(s_1,c) = -\frac{s_1}{2} \pm \frac{s_1}{2} \sqrt{ 1 + \frac{4c}{s_1^3} }.
\label{eq:lcsdr_s2}
\end{equation}
This is precisely the argument of the absorptive part worked out in \cite{song} (see  eq.~5 therein). The LCSDR can be worked out from \eqref{eq:roytrick_csdr} by  replacing $x_0\to -c/a$, and then taking  the limit $a\to 0$, yielding   
\be
M(s_1,s_2)=M(0,0)+\frac{1}{\pi}\int_{8/3}^\infty  \left[\frac{(2\sigma^3 -s_1 s_2 \left(s_1+s_2\right)) A(\sigma,s_2^{+}(\sigma,c))}{\sigma  \left(\sigma -s_1 \right) \left(\sigma -s_2\right) \left(\sigma +s_1+s_2\right)}-\frac{2}{\sigma}A(\sigma,0)\right]d\sigma,
\ee
We next discuss generalizations of CSDR on non-linear leaves. 

\section{Dispersion relations on nonlinear leaves}
\label{sec:nonlinear_leaves}

In this section, we consider more general foliations, not necessarily linear, defined by
\begin{equation}
y = f_\alpha(x),
\label{foliation}
\end{equation}
where \( f_\alpha \) is a function parameterized by \( \alpha \). Rather than attempting a complete classification, we focus on a few representative classes of phenomenologically relevant foliations, including an example already mentioned in the literature.

As noted in the Introduction, the motivation to consider nonlinear foliations stems from efforts to derive a generalized form of Roy equations that are valid across all energies and compatible with the rigorous analytic domain established by Martin.

To illustrate how the results of the previous section extend, let us suppose that the foliation is defined by a polynomial function \( f_\alpha(x) \). By restricting the amplitude to a given leaf, we can apply Cauchy's theorem around a point \( x \) where the function \( \mathcal{M}_\alpha(x) \equiv \mathcal{M}(x, f_\alpha(x)) \) is analytic:
\begin{equation}
\mathcal{M}_\alpha(x) = 
\frac{1}{2\pi i} \oint_x dx' \, \frac{\mathcal{P}(x') \mathcal{M}_\alpha(x')}{x' - x},
\label{cauchy_x}
\end{equation}
where \( \mathcal{P}(x') \) is a regulator function that suppresses the contribution from the arc at infinity.~\footnote{In general, nonlinear foliations may require additional subtractions to ensure convergence.}

Assuming the amplitude has no poles and the only singularity arises from the two-particle threshold, the location of the branch point on the leaf is obtained by solving
\begin{equation}
f_\alpha(x) = \frac{8}{3} \left(x - \frac{64}{9}\right).
\end{equation}
If \( f_\alpha(x) \) is a polynomial of degree \( N \), then generically this equation admits \( N \) solutions in the complex \( x \)-plane, corresponding to \( N \) threshold images, each associated with a branch cut to be integrated over. 
Then, the generalization of the dispersion relation \eqref{eq:disp_on_linear_leaf} to this case reads:
\begin{equation}
\mathcal{M}_\alpha(x) = \mathcal{M}_\alpha(x_s) + \sum_{i=1}^N \frac{1}{\pi} \int_{\mathcal{I}_i} dx' \, \frac{x - x_s}{x' - x_s} \, \frac{\mathcal{A}^i_\alpha(x')}{x' - x},
\label{eq:disp_on_non_linear_leaf}
\end{equation}
where \( \mathcal{A}^i_\alpha(x) \) denotes the discontinuity of the amplitude across the \( i \)-th cut \( \mathcal{I}_i \). 

If \( f_\alpha(x) \) is not a polynomial, additional singularities may be introduced by the foliation itself. In such cases, Eq.~\eqref{eq:disp_on_non_linear_leaf} must be modified to account for these extra features.

Analogous to the linear case, we start by solving the following equation for $x$
\begin{equation}
    P_\alpha(\sigma) = \sigma^3 - x \sigma + f_\alpha(x) 
    = 0 \, .
    \label{eq:sigma_multi}
\end{equation}

There exists in general $N$ solutions to \eqref{eq:sigma_multi}, which we denote as $x^{(i)}(\sigma)$ where $i \in \{1 \dots N\}$. Identifying $\sigma=s_1$ as before, we can obtain the solution for $s_2$ as follows
\be
    \tau^{(i)}(\sigma) = -\frac{\sigma}{2} + \frac{1}{2} \sqrt{4x^{(i)}(\sigma) - 3\sigma^2} \, .
    \label{eq:tau_multi}
\ee
Next, we need the Jacobian to perform the change of variables from $x'$ to $\sigma$ in \eqref{eq:disp_on_non_linear_leaf}. This can be obtained immediately by taking the $\sigma$-derivative of $P_\alpha(\sigma)$ yielding
\be
\frac{dx^{(i)}(\sigma)}{d\sigma} = \frac{3\sigma^2 -x}{\sigma-df/dx} = \frac{(\sigma-\tau^{(i)})(2\sigma+\tau^{(i)})}{\sigma-df/dx}\bigg|_{x=x^{(i)}} \, .
\ee
All in all, the pull-back of the dispersion relation \reef{eq:disp_on_non_linear_leaf} reads
\be
\boxed{
\begin{aligned}
M(s_1,s_2) - & M(s_1^0,s_2^0) = \sum_{i=1}^N \frac{1}{\pi}
\int^\infty_{8/3} \, d\sigma \,
\frac{dx^{(i)}(\sigma)}{d\sigma} \, A(\sigma,\tau^{(i)}(\sigma)) \\
&\times \left[ \frac{1}{x^{(i)}(\sigma) - s_1^2 - s_1 s_2 - s_2^2 } - \frac{1}{x^{(i)}(\sigma)- (s^0_1)^2 - s^0_1 s^0_2 - (s^0_2)^2 } \right]
\label{eq:pulled_disp_on_non_linear_leaf}
\end{aligned}
}
\ee
When $f_\alpha(x)$ is a linear function, the sum over $i$ contains a single term and  \eqref{eq:pulled_disp_on_non_linear_leaf} reduces to \eqref{eq:final_csdr} as expected. Note that the kernel in the above representation is manifestly crossing-symmetric.

Next we discuss few special cases. 

\begin{figure}[!t]
    \centering
    \includegraphics[width=.9\linewidth]{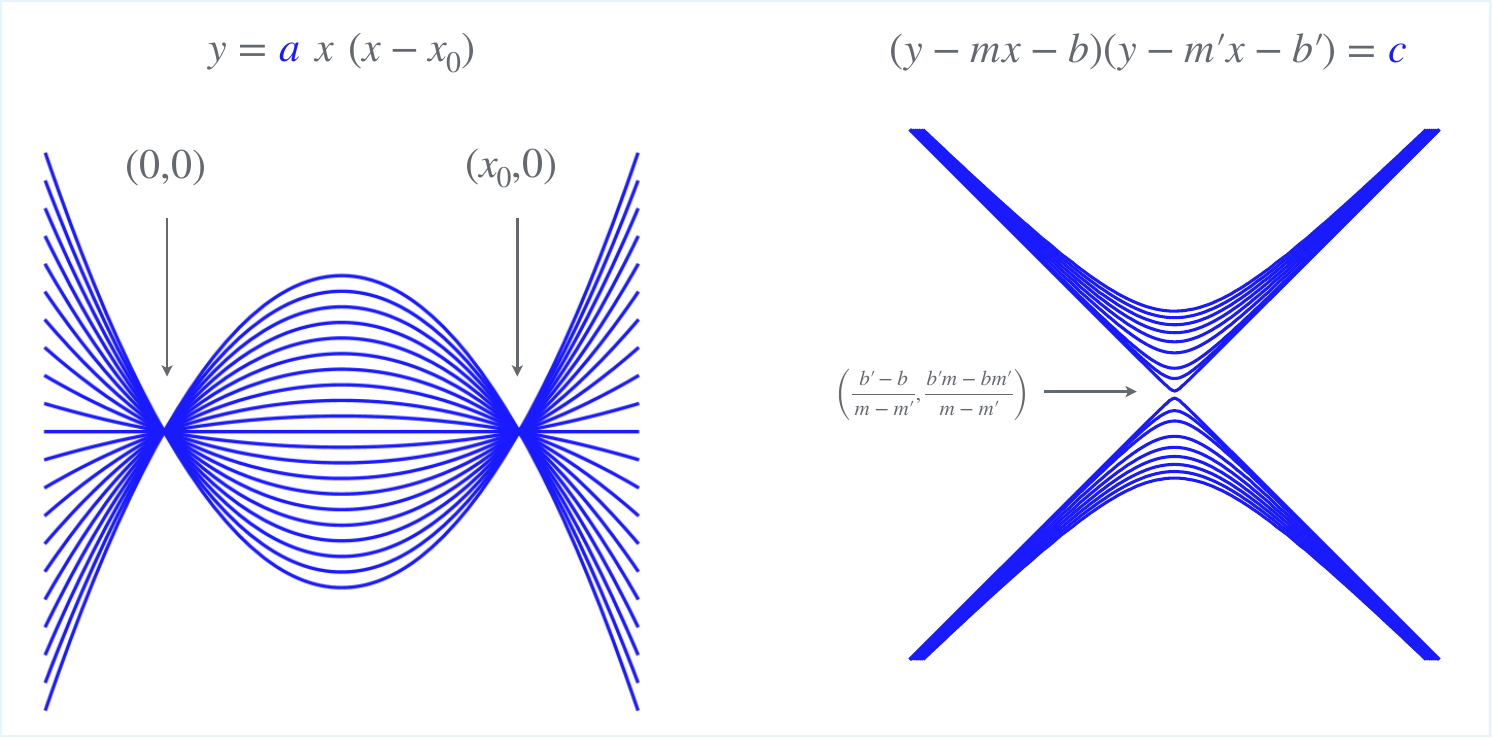}
    \caption{Foliations of the \( (x,y) \) plane by two conic sections. \textbf{Left:} Parabola foliation at fixed $x_0$ and various choices of $a$. \textbf{Right:} Hyperbola foliation at fixed $m,m',b,b'$ and various choices of $c$. 
    The points where all leaves intersect are also given in the figure -- see Section~\ref{nullc} for more details on properties of such points.
     }
    \label{fig:quad_foliations}
\end{figure}

\paragraph{A polynomial foliation.} 

To illustrate how dispersions on non-linear foliations work, let us consider the   parabola
\be
f(x)=ax(x-x_0).
\ee
At fixed $x_0$, this defines a foliation of $\mathbb{C}^2_{(x,y)}$ as we vary $a$. Its real projection can be seen in the left panel of Figure~\ref{fig:quad_foliations}.
The two solutions to \reef{eq:sigma_multi} are given by
\be
x^{(i)}(\sigma) =  \frac{(-)^i\sqrt{\left(a x_0+\sigma \right){}^2-4 a \sigma ^3}+a x_0+\sigma }{2 a} \quad , \quad i \in \{1,2\} \, ,
\ee
which, after combining with $\tau^{(i)}$ of \eqref{eq:tau_multi}, can be readily evaluated in \eqref{eq:pulled_disp_on_non_linear_leaf} to obtain the CSDR.

\paragraph{Auberson-Epele proposal.} Ref.~\cite{Auberson:1974in} proposed the hyperbola foliation 
\be
    (y-m ~x-b)(y-m'~x -b')=c \, ,
    \label{eq:auberson-epele}
\ee
without giving the explicit form of CSDR.
This constitutes one of the earlier attempts to extend the validity of the Roy-like equations using a curve which is non-linear. 
The parameters $m, m',b,b',c$ were chosen such that the amplitude is evaluated inside the rigorous Martin-Lehmann analytic domain~\eqref{ellipses}, whose geometry we extensively discuss in Section~\ref{roystuff}. It turns out that the validity of the Roy-like equations can be extended up to $s=164.4m^2$ for some choices. The leaves for various $c$ are illustrated in the right panel of Figure~\ref{fig:quad_foliations}.

Let us remark that both examples given above are two special cases of a generic conic curve foliating $\mathbb{C}^2_{(x,y)}$. Any other such choice -- an ellipse, parabola or hyperbola -- generally gives rise to two solutions for $x^{(i)}$, from which one can compute the respective CSDRs using \reef{eq:pulled_disp_on_non_linear_leaf}.

\paragraph{Homogeneous foliation.}
\label{sec:homogeneous}
Next we introduce a foliation that, to our knowledge, has not been discussed before,
\be
y=\alpha \, x^{3/2}.
\label{eq:homogeneous_foliation}
\ee
We term it homogeneous because  $\alpha$ is dimensionless. 
This leads to a CSDR that will find interesting applications in the context of extending Roy-like equations to all energies, as discussed in Sec.~\ref{roystuff} in more detail.

\begin{figure}[t!]
    \centering
    \includegraphics[width=\linewidth]{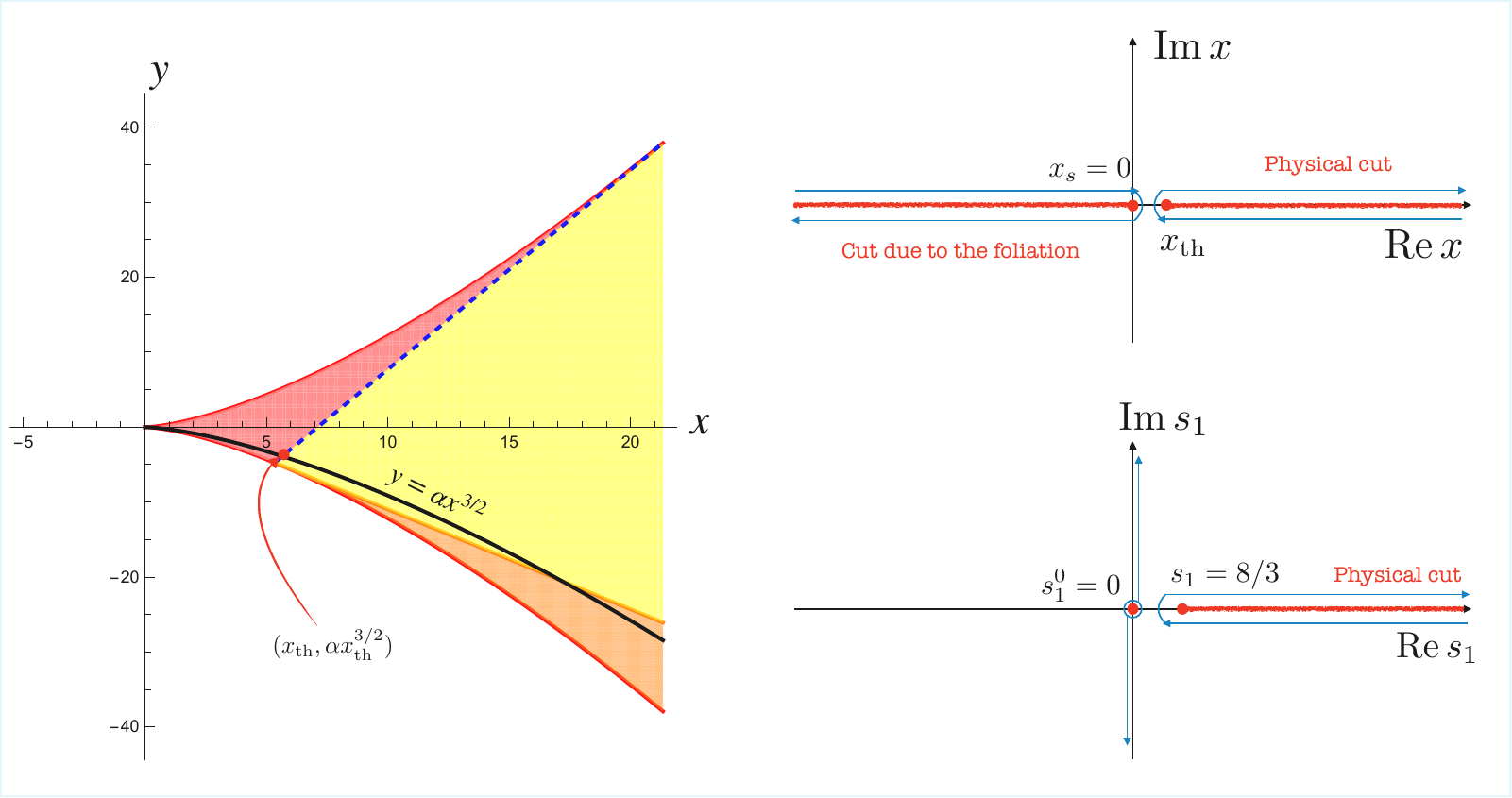}
    \caption{
    \textbf{Left:} A leaf of a non-analytic foliation \( f(x) = \alpha x^{3/2} \) in the \( (x, y) \) plane. When the leaf intersects the threshold line $y = \tfrac{8}{3}\left( x-\tfrac{64}{9} \right)$ (dashed blue), the amplitude develops a branch cut starting at $x_\text{th}$. \textbf{Right top:} Complex \( x \)-plane showing the deformed Cauchy contour. We have an additional left branch cut due to the foliation. \textbf{Right bottom:} The same Cauchy contour shown in the complex $s_1$-plane.
}
    \label{fig:foliation_x32}
\end{figure}

A key difference of this choice with respect to previous non-linear examples is that, it introduces additional singularities in the $x$-plane due to its fractional power, in the form of a branch cut starting at $(x,y)=(0,0)$ -- see Figure~\ref{fig:foliation_x32}. Moreover, it can cover the real wedge in the $x,y$ plane fully for the range $|\alpha| \leq \tfrac{2}{3\sqrt{3}}$.

When $\alpha\leq 0$, we can cover the physical scattering domain, and the leaf intersects the discontinuity region of the amplitude only once. Then, we can write
\be
\mathcal{M}_\alpha(x)=\mathcal{M}_{\alpha}(x_s) + \frac{1}{\pi} \int_{x_\text{th}}^\infty dx' \, \frac{x - x_s}{x' - x_s} \, \frac{\mathcal{A}_{\alpha}(x')}{x' - x}+\frac{1}{\pi }\int_{-\infty}^0 dx^\prime \frac{x - x_s}{x' - x_s} \, \frac{\text{Disc}_{x^\prime}\,{M}_{\alpha}(x')}{x' - x}.
\label{eq:nonlinear_in_x}
\ee

When pulling back, if we choose as subtraction point $x_s=0$ which is common to all leaves, we obtain a remarkable simplification of the kernel which gives the expression 
\begin{equation}
\boxed{
\begin{aligned}
&M(s_1,s_2)=M(0,0)+\\
&\frac{1}{\pi}\int_{8/3}^\infty \!\!\!\! d\sigma A(\sigma,\tau_\alpha(\sigma))\left(\frac{1}{\sigma-s_1}+\frac{1}{\sigma+s_1}-\frac{2}{\sigma}\right)+\frac{1}{\pi}\int_0^\infty \!\!\!\! d\sigma B_\alpha(\sigma)\left(\frac{i}{s_1-i\sigma}-\frac{i}{s_1+i\sigma}+\frac{2}{\sigma}\right),
\end{aligned}
}
\label{eq:nonanalytic_csdr}
\end{equation}
where $\tau_\alpha(\sigma) = -\tfrac{\sigma}{2} + \tfrac{1}{2} \sqrt{4x_\alpha(\sigma) - 3\sigma^2}$ with $x_\alpha(\sigma)$ being the unique solution to \eqref{eq:sigma_multi} for $\alpha <0$ -- see Appendix~\ref{details_roy_all} for its full form, and 
\be
B_\alpha(s)=
\frac{1}{2i}\left(M(i s,\tau_\alpha(i s))-M(-i s,\tau_\alpha(-i s))\right)=\im M(is,\tau_\alpha(is)),
\ee
where we used the fact that $\tau_\alpha(-is)=\tau_\alpha(is)^*$ (for $\alpha<0$), and by real analyticity $M(s_1^*,s_2^*)=M(s_1,s_2)^*$.
When mapping back from $x$ to $s$, we have to carefully keep track of where the negative $x\pm i \epsilon$ is mapped in the $s$-plane. The second term in the dispersion is equivalent to integrating the amplitude over the positive imaginary axis in the $s_1$ plane.

We remark that the equation~\reef{eq:nonlinear_in_x} was manifestly crossing-symmetric since the kernel depends on the $x$ variable. After doing the pull-back, the kernel in~\reef{eq:nonanalytic_csdr} is still crossing-symmetric though not manifest, because the relation between $s_1$ and $s_2$ is fixed via $x$ and $\alpha$ to stay on the same leaf in a non-trivial way. It is worth noting that there is a non trivial simplification happening when we express $\alpha$ in terms of the physical Mandelstam variables
\be
\tau_\alpha(\sigma)=\frac{s_2}{s_1}\sigma.
\ee
As we will see later, this identity simplifies drastically the partial wave projections of \eqref{eq:nonanalytic_csdr}. 

\section{Null constraints and singularities of foliations}
\label{nullc}

Null constraints are an infinite set of linear sum rules satisfied by the imaginary parts of the partial waves which take the form:
\begin{equation}
\int_4^\infty \!\!\!\! ds \sum_{\ell=0}^\infty \mathrm{Im}\, f_\ell(s)\,F_\ell^p(s) = 0,
\label{nullcons}
\end{equation}
where $F_\ell^p(s)$ is a known kernel function depending on a set of parameters \( p \).
These types of constraints have appeared in the phenomenological literature in the 1970s~\cite{Roskies:1970uj, Roy:1971tc}, and have been used to constrain solutions to the Roy equations~\cite{Ananthanarayan:2000ht}. 
More recently, they have been revived and extended in a modern dispersive bootstrap context \cite{Tolley:2020gtv,Caron-Huot:2020cmc}, and applied to constrain Wilson coefficients in weakly coupled EFTs and large-$N$ theories \cite{Chowdhury:2021ynh,Bhat:2023puy,Albert:2022oes,Albert:2023jtd,Albert:2023seb,Albert:2024yap,Bern:2021ppb,Caron-Huot:2022ugt,Fernandez:2022kzi,Ma:2023vgc,Haring:2023zwu,McPeak:2023wmq,Dong:2024omo,
Berman:2023jys,Berman:2024eid,Berman:2024wyt,Berman:2025owb%
}.  

In \cite{EliasMiro:2022xaa}, null constraints were explicitly derived from crossing symmetry via fixed-$t$ dispersion relations, yielding an expression reminiscent of the conformal bootstrap equation:
\begin{equation}
M(s,t) - M(s,4-s-t) = 0.
\end{equation}
Taking derivatives of this equation with respect to both \( s \) and \( t \) at a given kinematic point, and plugging in the partial wave expansion of the dispersive representation for the amplitude, leads to the constraints as in \eqref{nullcons}.

In the context of CSDRs, null constraints emerge more subtly: they appear as “locality conditions”~\cite{Sinha:2020win}. This naturally raises the question: if CSDRs are manifestly crossing-symmetric, why are null constraints not automatically satisfied? And more generally, what does locality mean in this setting? We address these issues in three steps.

\begin{itemize}
    \item CSDRs define analytic functions on each leaf of a foliation.
    \item Spurious poles or non-analytic behavior arise when expanding around \textit{singular points} of the foliation, as will be defined below.
    \item Enforcing analyticity at the singularities leads directly to null constraints.
\end{itemize}

With this argument, we conclude that in general null constraints are an intrinsic property of fixed-variable dispersion relations, and they arise as compatibility conditions between the geometric foliation of the kinematical space and the analytic structure of the amplitude. 

The theory of foliations of $\mathbb{C}^2$ and their singularities is a vast subject by itself \cite{laurentgengoux2024invitationsingularfoliations}, and our goal is not to discuss it in full generality. Instead, we will focus on the special class of foliations whose singularities are isolated and that can be understood with a simple local analysis.
The short section that follows contains only the information that is needed to understand the connection between foliation singularities and null constraints.

\subsubsection*{Local analysis and singular points}

\begin{figure}[t!]
    \centering
    \includegraphics[width=\linewidth]{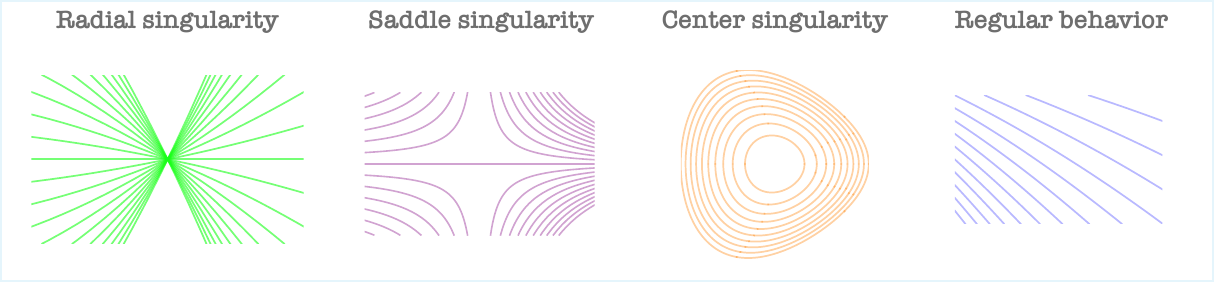}
    \caption{
    Locally, we can find either  three types of singularities, or a locally regular foliation.
    Note that the center and saddle singularities are only different real sections of a complex saddle point.
    }
    \label{fig:foliation_singularities}
\end{figure}

Intuitively, a foliation decomposes $\mathbb{C}^2$ into one-dimensional complex sub-manifolds, the \emph{leaves}, which locally resemble families of parallel lines. In our setting, a foliation can be specified by a holomorphic vector field. The leaves are then the integral curves of this vector field.

Let us illustrate this in a simple example. Consider the foliation used in the AK dispersion relations~\cite{Auberson:1972prg}, given by the line $y=ax$ for various values of the slope $a$. The level sets for constant slopes are defined by
\be
F(x,y) \equiv \frac{y}{x}=\text{const}.
\ee
The associated tangent vector field $\mathbf{T}(x,y)$ must be orthogonal everywhere to the normal vectors of the level sets given by the gradient of $F$, namely $\mathbf{T} \cdot \vec{\nabla} F=0$. In this case,
\be
\vec{\nabla} F = 
\begin{pmatrix} -y/x^2 \\  1/x \end{pmatrix} \quad\implies\quad \mathbf{T}(x,y) = \begin{pmatrix} x \\ y \end{pmatrix}  .
\label{eq:simple_vec_field}
\ee
We can expand this vector field  around a generic point $(\bar x,\bar y) \in \mathbb{R}^2$, and we obtain
\[
\mathbf{T}(x,y)=\mathbf{T}(\bar x,\bar y) + \mathbf{J}(\bar x,\bar y)\cdot 
\begin{pmatrix}x-\bar x\\ y-\bar y\end{pmatrix}+\dots,
\]
where $\mathbf{J}(\bar{x},\bar{y}) \equiv
    \begin{pmatrix}
    \partial_x \mathbf{T} \quad \partial_y \mathbf{T}
    \end{pmatrix}$ is the Jacobian matrix. For our example in~\reef{eq:simple_vec_field}, we have
\[
\begin{pmatrix}x\\y\end{pmatrix}
=
\begin{pmatrix}\bar x\\ \bar y\end{pmatrix}+
\begin{pmatrix}1&0\\0&1\end{pmatrix}
\begin{pmatrix}x-\bar x\\y-\bar y\end{pmatrix}+\dots
\]
Notice that, when $\bar x,\bar y\neq 0$, the zeroth-order approximation to the $\mathbf{T}$ above is the constant vector $(\bar x,\bar y)^T$, which is an example to a \emph{regular point}. In a neighborhood of a regular point $(\bar{x},\bar{y})$, the foliation looks like a family of parallel lines tangent to the constant vector $\mathbf{T}(\bar{x},\bar{y})$.

On the other hand, a point where vector field vanishes, $\mathbf{T}(\bar x,\bar y)=0$, is called a \emph{singular point}. In this example, it only occurs at $(0,0)$. The local behavior is now dictated by the Jacobian: here $\mathbf{T}(x,y)\simeq(x,y)$. This can be interpreted as a linear system of differential equations $\{ \dot x=x,\ \dot y=y \}$, whose solutions yield the level sets. The particular solutions in this case are radial lines through the origin, showing that near the singular point $(0,0)$ the foliation cannot be approximated by parallel lines.

In general, the types of singularities are determined by the Jacobian. When the real parts of the eigenvalues of $\mathbf{J}$ both carry the same sign, we get a \emph{radial} singularity; when the signs are opposite, we get a \emph{saddle} point. In Table~\ref{tab:foliations}, we summarized the local analysis for the examples introduced earlier. Linear, parabolic, and homogeneous foliations all display radial singularities. For hyperbolic or circular foliations, a holomorphic change of variables (e.g. $u=y-mx-b,\ v=y-m'x-b'$ or $u=(x-x_0)+i(y-y_0),\ v=(x-x_0)-i(y-y_0)$) reduces the description to the canonical form $uv=\text{const}$, corresponding to a complex saddle.

\begin{table}[t!]
    \centering
    \renewcommand{\arraystretch}{1.4} 
    \setlength{\tabcolsep}{10pt}      
    \begin{tabular}{c|c|c|c}
       \textbf{Level sets $F(x,y)$} & \textbf{Vector field $\mathbf{T}(x,y)$} & \textbf{Singularities} & \textbf{Jacobian $\mathbf{J}(x,y)$}  \\
       \hline
       $\dfrac{y}{x-x_0}$  & 
       $\begin{pmatrix} x-x_0 \\[2pt] y \end{pmatrix}$ & 
       $(x_0,0)$ & 
       $\begin{pmatrix} 1 & 0 \\[2pt] 0 & 1 \end{pmatrix}$ \\ 
       \hline
       $\dfrac{y}{x(x-x_0)}$ & 
       $\begin{pmatrix} x(x-x_0) \\[2pt] y(2x-x_0) \end{pmatrix}$ & 
       $\{(0,0),(x_0,0)\}$ &
       $\begin{pmatrix} 2x-x_0 & 0 \\[2pt] 2y & 2x-x_0 \end{pmatrix}$ \\ 
       \hline
       $uv$ & 
       $\begin{pmatrix} u \\[2pt] -v \end{pmatrix}$ &
       $u=v=0$ & 
       $\begin{pmatrix} 1 & 0 \\[2pt] 0 & -1 \end{pmatrix}$ \\ 
       \hline
       $\dfrac{y^2}{x^3}$ & 
       $\begin{pmatrix} 2x \\[2pt] 3y \end{pmatrix}$ & 
       $(0,0)$ & 
       $\begin{pmatrix} 2 & 0 \\[2pt] 0 & 3 \end{pmatrix}$ \\ 
    \end{tabular}
    \caption{Examples of foliations defined by their level sets, tangent vector fields, singular sets, and Jacobians.}
    \label{tab:foliations}
\end{table}

\subsubsection*{Compatibility conditions at singular points}

After the classification of possible singularities, we now turn to the analytic properties of $\mathcal{M}(x,y)$ at such points.


\paragraph{Analytic continuation and foliation singularities.}

Naturally, if the absorptive part of the amplitude 
$\mathcal{A}(\sigma,\tau(\sigma))$ on each leaf of $y=f_\alpha(x)$ is integrable, 
the amplitude is analytic on the leaf except for the physical singularities 
or those introduced by the foliation itself, as in the homogeneous case.

Extending the amplitude to a function of \emph{two} complex variables 
by promoting the foliation parameter $\alpha$ to a complex quantity requires 
additional care. To clarify this connection, expand the absorptive part 
in partial waves of the physical scattering angle $\theta$,
\begin{equation}
A(\sigma,\tau^{(i)}(\sigma))
= \sum_{\ell=0}^{\infty}
n_\ell\,\mathrm{Im}\,f_\ell(\sigma)\,P_\ell(\cos\theta),
\label{eq:pw_expansion}
\end{equation}
where \(i\) labels possible multiple discontinuities induced by a nonlinear foliation as discussed in Section~\ref{sec:nonlinear_leaves}.
The cosine of the scattering angle (denoted \(\sqrt{\xi}\) in the CSDR literature~\cite{Sinha:2020win}) 
depends on the foliation parameters via \(\tau^{(i)}\) or \(x^{(i)}\) defined in~\eqref{eq:tau_multi}:
\begin{equation}
\cos\theta =
\frac{\sigma+2\tau^{(i)}(\sigma)}{\sigma-\tfrac{8}{3}}
= \frac{\sqrt{4x^{(i)}(\sigma)-3\sigma^2}}{\sigma-\tfrac{8}{3}},
\end{equation}
which is analytic in the complex \(\cos\theta\) plane inside an elliptic domain terminating 
at the first singularity -- see Section~\ref{sec:elliptic_domains} for further details.

Even within the analytic domain, expanding \(\cos\theta\) in the foliation 
parameters can impose \emph{non-trivial consistency conditions} that may 
clash with the leaf structure. To illustrate, consider first a regular point 
of a foliation where the leaves are approximately parallel and locally 
approximated by \(y=a(x-x_0)\). Using~\eqref{eq:x_match}, the Legendre 
argument becomes\footnote{Only even spins contribute because of crossing symmetry.}
\begin{equation}
\cos^2\theta =
\frac{\sigma^3 + 3a\sigma^2 - 4(a x - y)}
{(\sigma-a)(\sigma-\tfrac{8}{3})^2},
\label{eq:xi_def}
\end{equation}
which is analytic in \(x,y\). Varying \((x,y)\) at fixed \(a\) shifts 
\(x_0 = x - y/a\) across parallel leaves, leaving analyticity intact.

By contrast, near a \emph{singular point}, for instance a radial singularity 
common in the literature, the foliation is approximated by 
\(y=a(x-x_0)\) (with \(y_0=0\) for simplicity). 
The Legendre argument now reads
\begin{equation}
\cos^2\theta =
\frac{x_0(4y+\sigma^3) - x\sigma^3 - 3\sigma^2y}
{(\sigma-\tfrac{8}{3})^2 \,(y-\sigma(x-x_0))},
\label{eq:singular_angle}
\end{equation}
which becomes singular at \((x,y)=(x_0,0)\). Approaching 
\((x-x_0,y)\to0\) can generate arbitrarily high-order poles. 
Avoiding spurious singularities therefore requires additional regularity 
conditions on the amplitude, the \emph{null constraints} in this 
geometric setting.

A similar analysis applies to saddle singularities. Explicit derivations 
for linear and quadratic foliations are provided in 
Appendix~\ref{compcond}. Moreover, even for linear regular foliations 
with parallel leaves \(y=c\), expanding at infinity leads to analogous 
sum rules, implying that the local CSDRs~\cite{song} must also 
be supplemented by null constraints.

\section{Extending Roy-like equations to all energies}
\label{roystuff}

Thus far, we have primarily discussed crossing symmetry and analyticity constraints on dispersion relations from a general standpoint. However, the motivation for introducing nonlinear foliations stems from a more specific goal: constructing dispersion relations for partial wave amplitudes that are valid at all energies and fully compatible with the rigorous analyticity domain established by Martin~\cite{Martin:1965jj,Martin:1966zsy}. 

Before proceeding in this direction, we take this opportunity to briefly review Martin’s result on analyticity. Rather than presenting the full technical details of the proof, we focus on adapting his conclusions to the framework relevant for our purposes.

\subsection{Analyticity Domains and Fixed-$t$ Dispersion Relations}
\label{sec:elliptic_domains}

Analyticity is a recurring theme in any S-matrix bootstrap discussion. For the scattering of the lightest particle in a theory, the most natural assumption is the principle of \emph{maximal analyticity}~\cite{Mandelstam:1958xc}: the amplitude is analytic in all Mandelstam variables, except along cuts associated with physical processes. This ``smart'' domain of analyticity introduced by Mandelstam has passed every perturbative test, yet continues to elude a general proof -- see~\cite{Correia:2020xtr} for a review and~\cite{Correia:2021etg} for an extended analysis of Landau singularities in generic diagrams. 
An interesting approach to prove maximal analyticity comes from studying the flat-space limit of conformal correlators~\cite{Paulos:2016fap}, and try to derive it as consequence of CFT axioms \cite{vanRees:2022zmr,vanRees:2023fcf}.
In this section, we restrict ourselves to the rigorous domain derived by Martin in~\cite{Martin:1965jj} and extended in~\cite{Martin:1969ina}. For the time being, we will switch back to the usual Mandelstam invariants $s,t,u$. 

Starting from the fixed-$t$ dispersive representation
\be
M(s,t)=\int_4^\infty A(s',t) K(s',s,t)+\text{subtractions}
\ee
Martin proved that the absorptive part of the amplitude $A(s,t)$ is analytic in $t$ for all $s>4$ whenever $t$ belongs to a complex domain $\mathcal{D}$ containing the real segment $-28 \leq t < 4$, thereafter showing the full amplitude $M(s,t)$ being analytic in the product set $\mathcal{D} \times (s-\text{plane minus cuts})$.
This domain is constructed as the intersection over all $s \geq 4$ of the union of the Lehmann and Martin ellipses:
\begin{equation}
\mathcal{D} = \bigcap_{4 \leq s < \infty} \left( E_L(r(s), s) \cup E_U(4, s) \right) = \bigcap_{16 \leq s \leq 64} E_L(r(s), s),
\label{ellipses}
\end{equation}
where the elliptic domain is defined by  
\be
   E(\bar{r},s)=\left\{ (s,t)\Big| (s,t)\in \frac{(\mathrm{Re}(t) - f_2/2)^2}{(\bar{r} - f_2/2)^2} + \frac{(\mathrm{Im}(t))^2}{\bar{r}(\bar{r} - f_2)} < 1 \right\}, \quad \text{with } f_2 = 4 - s.
    \label{eq:elip}
    \ee
The value of right extremity $\bar{r}$ depends on whether we are considering the  Martin or Lehmann ellipse. 
\begin{itemize}
  \item $E_U(R,s)$ is the Martin ellipse with foci at $t=0$ and $t = 4 - s$ and right extremity $R = 4$, in the absence of bound states;~\footnote{The presence of a spin-two bound state would shrink the right extremity to $t=m^2_b$.}
  \item $E_L(r(s), s)$ is the Lehmann ellipse with the same foci and right extremity $r(s) = 256/s$.
\end{itemize}

\begin{figure}[t!] 
\centering 
	\includegraphics[width=0.9\textwidth]{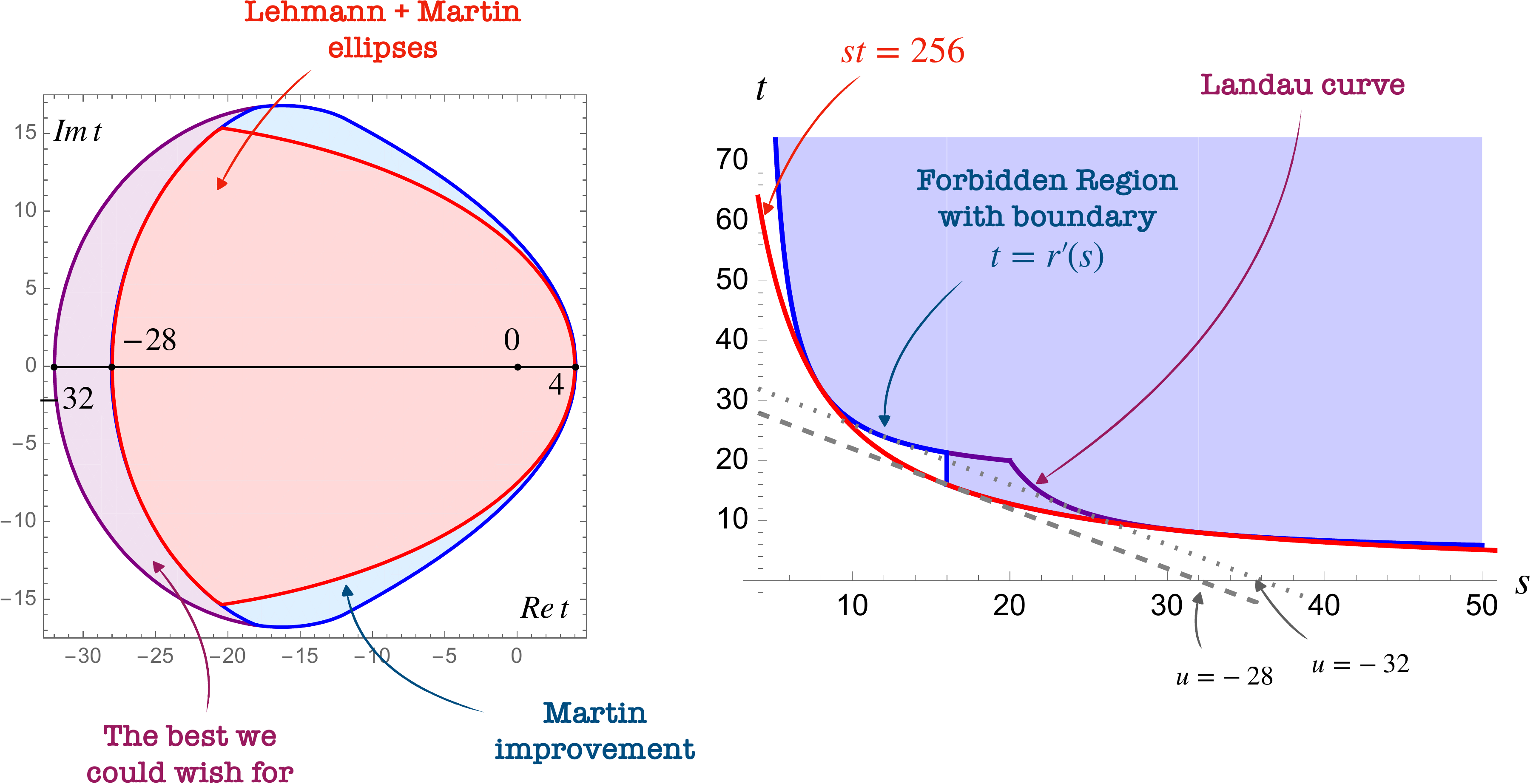} 
	\caption{\textbf{Left:} domain $\mathcal{D}$ (in red), extended domain $\mathcal{D}^\prime$ (in blue), and the domain we should wish for (in purple). \textbf{Right:} With the same color, the boundary of the analyticity region projected on the plane of real $s$ and $t$ variables.}
\label{ellipses_Martin}
\end{figure}

The crucial ingredient in the definition of $\mathcal{D}$ is the existence of the Martin ellipse.
Without the Martin ellipses, the intersection of all Lehmann ellipses would coincide with the real segment $-28 \leq t < 4$, since the ellipse becomes degenerate when $s\to \infty$. The right-hand side of~\eqref{ellipses} results from a simple geometric analysis. In Figure~\ref{ellipses_Martin} (left), we illustrate in red the domain $\mathcal{D}$ obtained from this construction.

This domain was further enlarged by Martin in~\cite{Martin:1969ina} -- see also the blue region in Figure \ref{ellipses_Martin} (left)
\begin{equation}
\mathcal{D}^\prime = \bigcap_{4 \leq s < \infty} E_L(r'(s), s),
\label{larger_ellipses}
\end{equation}
where the extremity function $r'(s)$ is given piecewise by~\footnote{Dear reader, if this piece-wise domain looks unnatural to you, you are not alone. 
We do not have a concise understanding of the origin of this piecewise function. Furthermore, it   would be nice to prove analyticity below the Karplus curve.}:
\begin{equation}
r'(s) =
\begin{cases}
\displaystyle \frac{16s}{s - 4}, & s \leq 16, \\[10pt]
\displaystyle \frac{256}{s}, & 16 \leq s \leq 32, \\[10pt]
\displaystyle \frac{4s}{s - 16}, & s \geq 32.
\end{cases}
\label{extremities}
\end{equation}

In Figure~\ref{ellipses_Martin} (right), we give an alternative view of the extremity of these ellipses projecting them on the plane of real $s$ and $t$ variables. The red curve corresponds to the Lehmann ellipse extremity $t=256/s$. The dashed gray line intersecting this curve is the line $u=\text{constant}=-28$. This gives an intuition where the number $-28$ comes from.

The blue curve represent the extremity of the improved domain found by Martin. Looking at its shape, it seems pretty unnatural. Indeed, 
 the purple curve is the unitary Landau curve, along which the amplitude’s discontinuity becomes singular. This curve seems the natural boundary for the validity of fixed-$t$ dispersion relations. To our knowledge, no rigorous extension of analyticity between the blue and purple curves has yet been achieved. 

 Assuming we were able to prove analyticity up to the Landau curve, we would be able to extend analyticity in $t$ to the domain depicted in purple in Figure \ref{ellipses_Martin} (left).

\subsection{The analyticity domain in $x,y$ variables}

For each fixed value of $s$, the domain of analyticity in the $t$-channel is bounded by an ellipse with foci at $(0,0)$ and $(4-s,0)$, and right extremity given by \eqref{extremities}. This ellipse can be parametrized by the complex function
\begin{equation}
t(s,\phi) = c(s) + (r'(s) - c(s)) \cos\phi + i \sqrt{(r'(s) - c(s))^2 - c(s)^2} \, \sin\phi,
\end{equation}
where $c(s) = 2 - s/2$ denotes the center of the ellipse.

The corresponding parametric curve is given by $\mathcal{C}(s,\phi) = (x(s,t(s,\phi)), y(s,t(s,\phi)))$ in the $(x,y)$ plane. This curve lies entirely in the real $(x,y)$ plane only when $\sin(2\phi) = 0$. Due to the crossing symmetry of the $x,y$ variables, only the values $\phi = 0$ and $\phi = \pi$ yield distinct curves.

\begin{figure}
    \centering
    \includegraphics[width=0.55\linewidth]{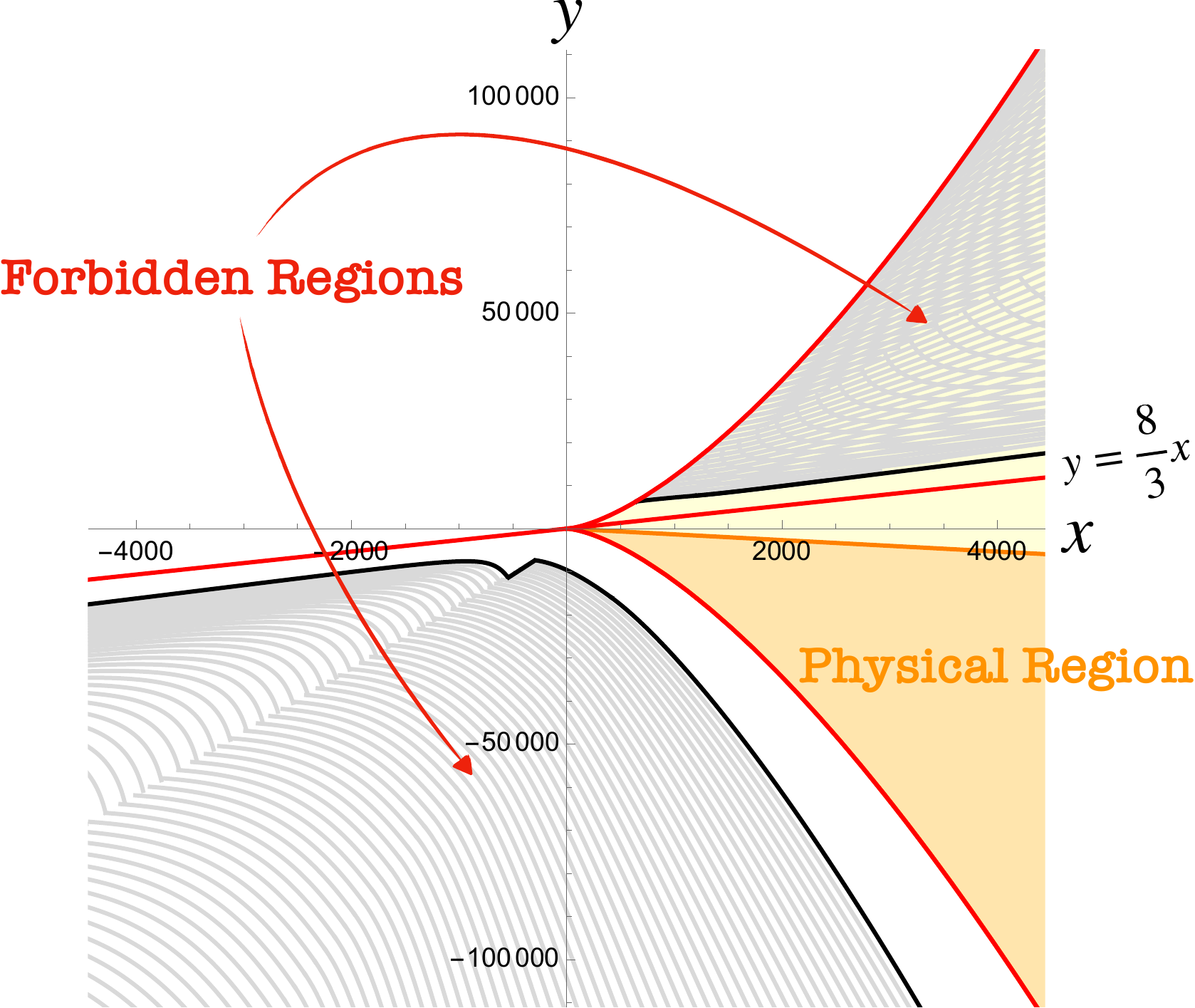}
    \caption{The analyticity domain in the $(x,y)$ plane. The orange region corresponds to the physical domain of real scattering angles and physical energies. The yellow region enveloping it is the real fundamental domain. Black lines denote the boundaries of the analyticity domain. The upper-right excluded region corresponds to the image of the double-discontinuity domain in $(x,y)$ space. The lower boundary reflects the upper extremity of the ellipse in complex $t$.}
    \label{fig:analyticity_in_x_y}
\end{figure}

To visualize the analyticity domain in $x,y$ variables, it is sufficient to plot these curves as a function of $s$, as shown in Figure~\ref{fig:analyticity_in_x_y}. The red region in orange represents the physical scattering domain; the surrounding yellow region is the maximal real domain compatible with analyticity. The black boundaries indicate regions where we haven't proven the amplitude is analytic.

Asymptotically, the forbidden region for real values of $(x,y)$ can be approximated by the following inequalities:
\begin{align}
&-\frac{2}{3\sqrt{3}}x^{3/2} \leq y \leq \frac{8}{3}x \quad \text{or} \quad y \geq \frac{2}{3\sqrt{3}}x^{3/2}, \quad \text{for } x \geq 0, \nonumber \\
&y \geq \frac{8}{3}x, \quad \text{for } x \leq 0.\label{eq:asymptotic_domain}
\end{align}

A necessary condition for an admissible foliation that fully covers the physical region is that it remains within this approximate asymptotic domain. Of course, once such a foliation is constructed, one must also verify that it remains inside the full analyticity domain for complex values of $(x,y)$. On the other hand, if one is interested only in covering a finite portion of the domain, it becomes important to take into account the detailed shape of the analyticity region in $(x,y)$ space.

\subsection{Partial wave projection in $x,y$ space}

To derive rigorous dispersion relations for partial waves, it is not enough to identify a single leaf that remains within the analyticity region; one must construct a family of leaves which, for each fixed energy $s$, covers the segment $2 - s/2 \leq t \leq 0$.

\begin{figure}
    \centering
    \includegraphics[width=0.6\linewidth]{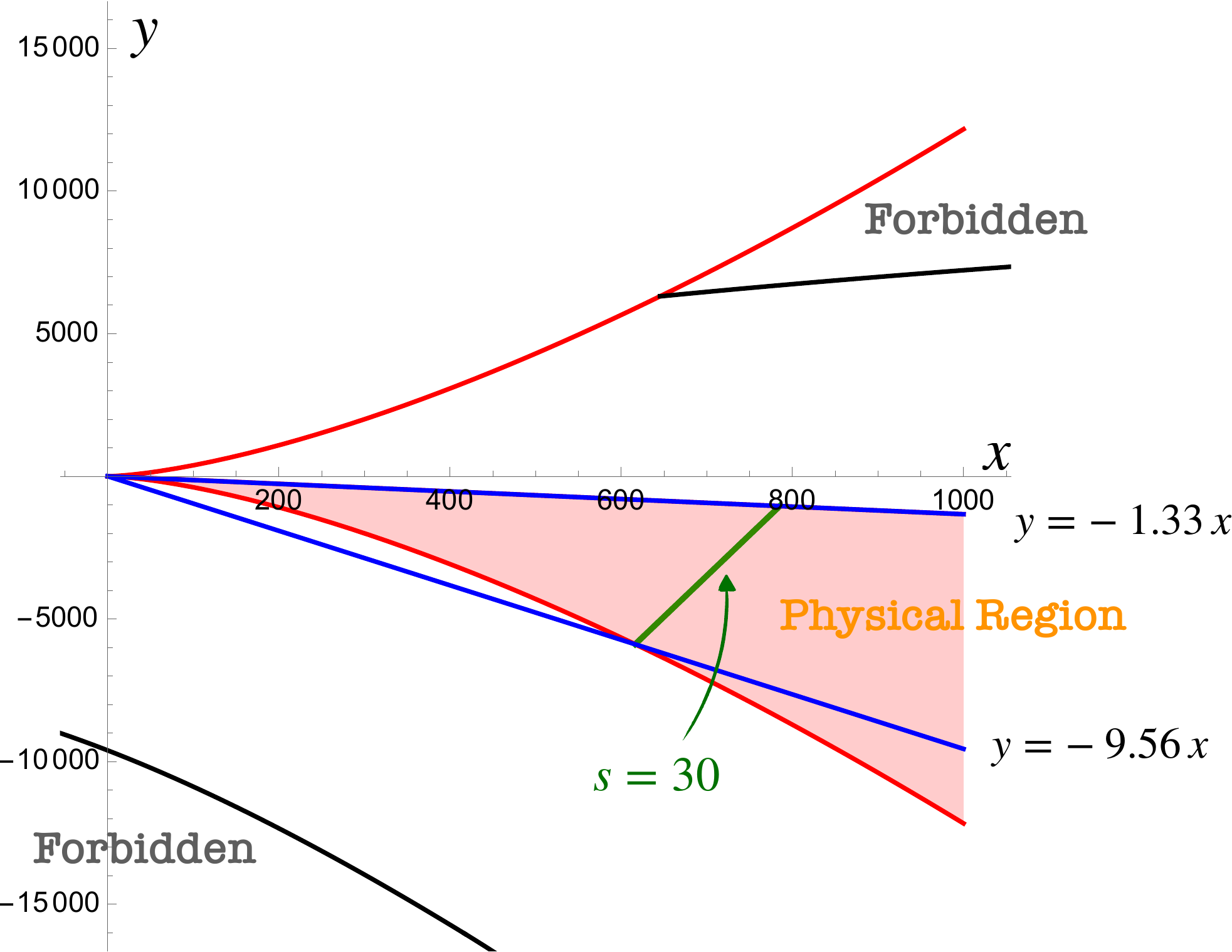}
    \caption{Projection segment for $s = 30$ (green), and an admissible foliation using $x_0 = 0$ covering it for $-9.56\leq a \leq -1.33$.}
    \label{fig:partial_wave_projection}
\end{figure}

Following \cite{Mahoux:1974ej}, we note that at fixed $s$, this segment maps in the $(x,y)$ plane to a straight line:
\begin{equation}
    y = s_2 x - s_2^3 = (t - 4/3)x - (t - 4/3)^3, \quad \text{with } 2 - s/2 \leq t \leq 0.
\end{equation}
At $t = 0$, this yields the line $y = \left(\tfrac{4}{3}\right)^3 - \tfrac{4}{3}x$, while $t = 2 - s/2$ corresponds to the lower boundary of the physical region, $y = -\tfrac{2}{3\sqrt{3}}x^{3/2}$.

Figure~\ref{fig:partial_wave_projection} illustrates the segment that must be integrated over to project onto partial waves at $s = 30$. When using a linear foliation $y = a(x - x_0)$ with $x_0 = 0$, we must ensure that the family of leaves spans this segment while remaining inside the analyticity domain. In this example, this is achieved for $-9.56 \leq a \leq -1.33$. 

More generally, for any given foliation, one must solve a geometric problem: find the range of parameters \((a, x_0)\) for which the leaves both remain analytic and allow projection to partial waves up to the largest possible $s$. This is the origin of the bounds reviewed in Figure~\ref{plot_cutoffs}.

\paragraph{Analyticity of the linear foliation $y = a(x - x_0)$.}

This analysis was carried out in \cite{Auberson:1972prg, Mahoux:1974ej} and can be approached geometrically. Figure~\ref{fig:analyticity_in_x_y} shows that not all straight lines lie within the analyticity region. The upper bound on admissible slopes is set asymptotically by the asymptotic Landau curve $y = \tfrac{8}{3}x$, so we must have $\mathrm{Re}(a) < \tfrac{8}{3}$. The lower bound arises from the point at which the foliation line touches the bottom boundary of the analytic domain.

\begin{figure}[t!] 
\centering 
	\includegraphics[width=1\textwidth]{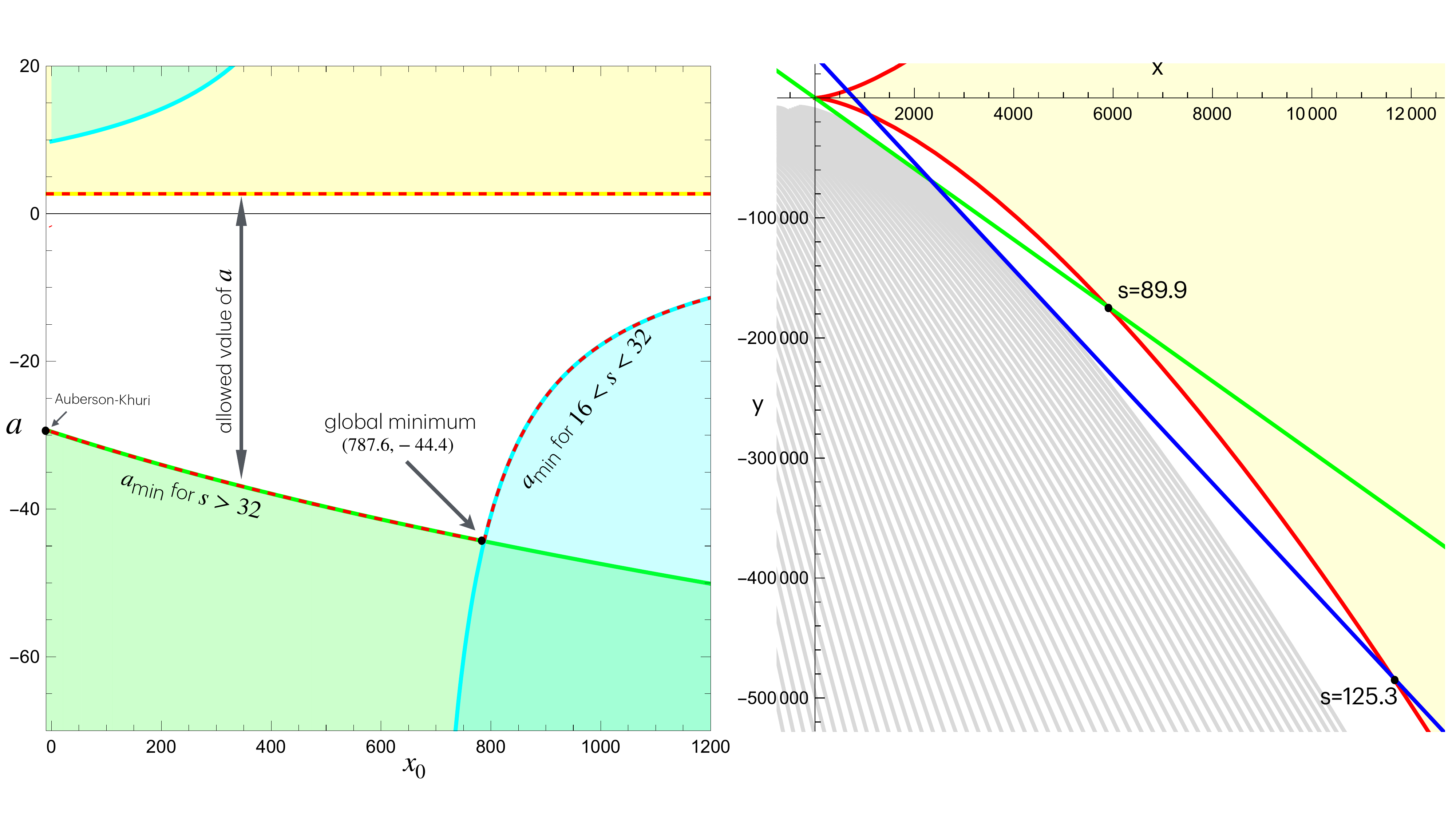} 
	\caption{\textbf{Left:} Allowed range of $a$ as a function of $x_0$. For $x_0 = 0$, the range is $-29.5 < a < 8/3$; the optimal choice is $x_0 = 787.6$, for which $-44.4 < a < 8/3$. \textbf{Right:} Gray region is excluded by analyticity; the yellow region is the real fundamental domain.}
	\label{ax0regions}
\end{figure}

Figure~\ref{ax0regions} (Right) illustrates this for $x_0 = 0$, where the minimal slope is $a_\text{min} = -29.5$, corresponding to a maximal projection energy of $s = 89.90$. Varying $x_0$ yields the family of admissible regions shown in Figure~\ref{ax0regions} (Left), where the optimal value is $x_0 = 787.6$, allowing projection up to $s = 125.31$ with $a_\text{min} = -44.4$.

A generic feature of linear foliations is that no single choice $(a, x_0)$ allows projection at all allowed energies. As already observed in \cite{Mahoux:1974ej}, covering the full energy range requires two separate dispersion relations:
 \begin{eqnarray}
 &&y=ax, \quad\quad\quad\quad\quad\,\, -29.5\leq a<8/3 \quad\quad\,\, 4\leq s\leq 89.90\nn\\ 
 &&y=a(x-787.6), \quad -44.4\leq a <8/3, \quad\quad 39.78\leq s\leq 125.31
 \end{eqnarray}
Each requires a distinct subtraction point: $(s,t) = (4/3, 4/3)$ for the first, and $(s,t) = (29.4, 4/3)$ for the second.

Polynomial foliations can in principle extend this range further, introducing multiple dispersion windows and subtraction constants. However, a unique case stands out: the non-polynomial foliation $y = \alpha x^{3/2}$, dictated by the asymptotic boundary \eqref{eq:asymptotic_domain}, already introduced in Section~\ref{sec:nonlinear_leaves}. This choice supports projection for all $s$ with a single family of leaves. The trade-off is the presence of a branch-point singularity at $(x,y) = (0,0)$, which introduces an extra contribution interpretable as a subtraction function.

\subsection{Homogeneous foliation $y=\alpha x^{3/2}$ }
\label{sec:homogeneous_roy}

For the homogeneous foliation introduced in Sec.~\ref{sec:homogeneous}, we determine the range of parameters needed for partial wave projection as
\be
    -\frac{2}{3\sqrt{3}}\leq \alpha \leq 0.
\label{eq:roy_infinity}
\ee
The reason for this choice is simple: In right panel of Figure~\ref{ax0regions}, the boundary of lower gray region asymptotically goes as $-\frac{2}{3\sqrt{3}}x^{3/2}$, therefore, $\alpha x^{3/2}$ curve with the restriction~\reef{eq:roy_infinity} avoids the Martin-Lehmann excluded regions for all physical energies and angles, allowing to project on partial waves for all energies (see more details in Appendix~\ref{details_roy_all}).

The corresponding set of Roy-like equations take the form
\be
\boxed{
\begin{aligned}
f_\ell(s_1)=&\frac{M(0,0)\delta_{\ell,0}}{16\pi}+\int_{\tfrac{8}{3}}^\infty d\sigma \sum_{\ell'=0}^\infty \im f_{\ell'}(\sigma)K^1_{\ell',\ell}(\sigma,s_1) \\
&+\int_0^\infty d\sigma \sum_{\ell'=0}^\infty \left[ \im f_{\ell'}(i\sigma) K^2_{\ell',\ell}(\sigma,s_1)+\re f_{\ell'}(i\sigma)K^3_{\ell',\ell}(\sigma,s_1) \right],
\end{aligned}
\label{royex}
}
\ee
where the detailed expressions for the kernels $K^i_{\ell',\ell}(\sigma,s_1)$ for $i=1,2,3$ 
can be found in Appendix~\ref{details_roy_all}, and the subtraction constant $M(0,0)$  is an unknown that, as usual in dispersion relations, needs to be provided.

Eq.~\eqref{royex} is obtained by partial wave projection of the dispersion relation \reef{eq:nonanalytic_csdr}. On the right-hand side, we make use of the partial wave expansion of the amplitude for physical and purely imaginary values for $\sigma$. We have shown in detail in Appendix~\ref{details_roy_all} that the expansion converges for both cases. 

\begin{figure}[t!] 
	\includegraphics[scale=0.75]{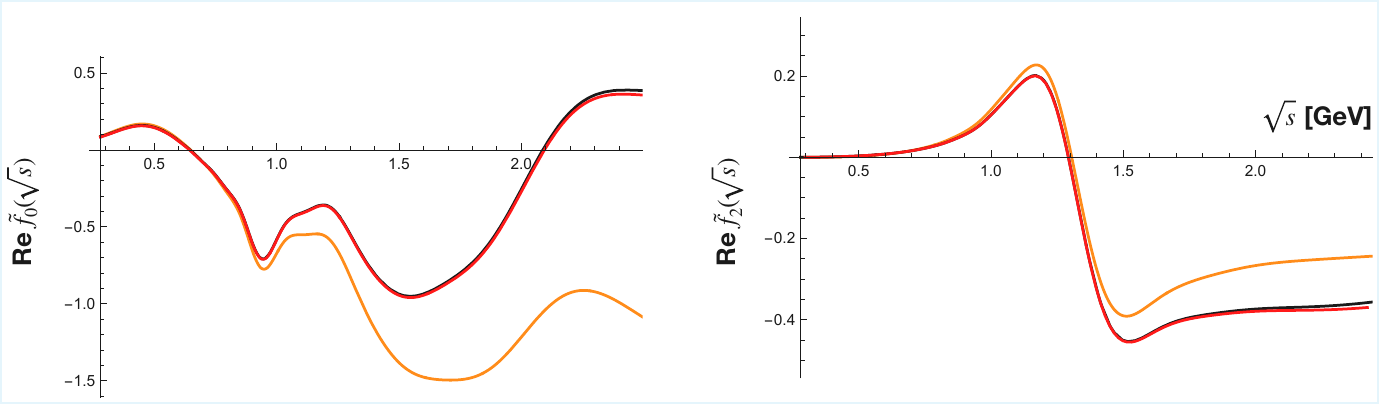} 
	\caption{Roy-like equations measured on the fully crossing-symmetric component of the non-perturbative pion amplitude $\tilde{M}(s,t,u)$. \textbf{Left:} Real part of $\tilde{f}_0(s)$ (in black) is compared against the right-hand side of~\reef{royex}: the first two terms (in orange) and all three terms (in red). \textbf{Right:} The same for $\tilde{f}_2(s)$.}
	\label{pion_amp_roy_like}
\end{figure}
 
The interesting aspect of this foliation is that such projection can be achieved for physical angles at all physical energies $s_1>8/3$, thus providing a Roy-like equation valid to all energies~\footnote{Note that any other foliation will always have an upper bound. For example, for AK-CSDR the restriction on parameter $a>-29.5$ gives upper bound $s<89.9$; and for MRW-CSDR, the restriction on parameter $a>-44.4$ gives upper bound $s<125.3$ (see Figure~\ref{ax0regions}). One can also show that, for constant foliation, the analogous restrictions, as explained in Appendix~\ref{sec:further_det_analyticity}, yields the upper bound $s<32.2$.}.
There is however a price to pay for this extension: one needs to provide 
$\text{Disc}_{x'<0}\mathcal{M}_\alpha(x^\prime)$ (last term in \eqref{royex}). 
Providing  $\text{Disc}_{x'<0}\mathcal{M}_\alpha(x^\prime)$ is  challenging because it is evaluated at nonphysical kinematics.
This seems to be a general feature of foliations that reach arbitrarily high energies—though a formal proof is lacking.

To test \eqref{royex}, we consider the non-perturbative pion amplitude constructed in \cite{Guerrieri:2024jkn}. In particular, we apply them to the crossing symmetric component $\tilde M(s,t,u)=\tfrac{1}{3}M^{(0)}(s,t,u)+\tfrac{2}{3}M^{(2)}(s,t,u)$, where $M^{(0)}$ is the singlet, and $M^{(2)}$ the symmetric traceless amplitude component.
In Figure~\ref{pion_amp_roy_like}, we plot in black the real part respectively of $\tilde f_0$ and $\tilde f_2$ for the first two spins as function of the energy in GeV as computed projecting $M^{(0)}$ in partial waves. Orange and red lines are respectively the contribution from the right-hand cut, and the sum of right-hand cut and imaginary line integral (the second line in \eqref{royex}). Interestingly, there is no contribution coming from spin zero to the spin two right-hand side integral. Moreover, the dispersion relations appear quite orthogonal since the difference between the true projection and our truncated reconstruction to a single wave is quite good up to 2.5 GeV.

\section{Outlook}

The old analytic S-matrix literature is filled with deep and often unresolved questions, which were largely abandoned due to technical difficulties and waning interest. The recent resurgence of interest in the non-perturbative S-matrix methods prompts us to revisit these problems with fresh tools and renewed motivation.

A fundamental open question concerns the Martin-Lehmann analyticity domain. 
We have now strong evidence \cite{Correia:2021etg} that the domain of the double-discontinuity is bounded by the two Karplus curves, which are simply given in the $s-t$ channel by the two hyperbolas $t=4s/(s-16)$ and $t=16s/(s-4)$. 
Interestingly, the Martin-Lehmann domain stops a little below these curves. It would be desirable to extend the proof of analyticity all the way to the double discontinuity, or otherwise prove the impossibility of such an extension and understand the reason for it.

The dispersion relations for partial waves \eqref{royex} that we proved are valid for all energies suggest that it should be possible to derive an extended analyticity region for partial wave amplitudes. It would be interesting to revisit the analysis done by A. Martin \cite{Martin:1965jj} in the light of this new set of dispersion relations.

On a more general ground, it is important to point that there is a conceptual gap between the analyticity domain derived by Bros, Epstein, and Glaser (BEG) — which relies purely on causality — and the Martin-Lehmann “potato,” which incorporates BEG analyticity together with positivity which we used to derive the results in this work. Could it be that using full non-perturbative unitarity (which is absent in Martin’s original analysis) could help bridging this gap?

A promising direction for future work is the study of dispersion relations for higher-point scattering amplitudes. While the proliferation of kinematic variables and crossed-channel singularities has historically made dispersion relations for such amplitudes technically daunting (see~\cite{Eden:1966dnq} for a classic reference on this subject), our framework may offer a new angle. In particular, it would be worth exploring whether multi-point dispersion relations can be formulated using crossing-symmetric variables. The main obstacle is the highly non-linear relation between these variables and the standard Mandelstam invariants, which complicates the connection to unitarity. To our knowledge, this direction remains largely unexplored, yet it could offer a new way to evade anomalous thresholds appearing in multi-particle amplitudes, which pollute dispersion relations written in terms of Mandelstam variables. 

In this work, we have found a set of Roy-like equation which are valid for all physical energies \eqref{royex}. 
The beauty of these equations is their simplicity. 
The foliations needed to derive them is highly non-trivial, but their final expressions are free of horrid cubic roots.
The resulting Roy equations are easy to write, and the resulting kernels are rational functions of the Mandelstam invariants. \footnote{The reader might ask why not derive these equations directly in the Mandelstam. That should be possible. Let us point out, however, that a priori would have been really hard to guess the integration contour in \eqref{eq:nonanalytic_csdr}.}
However, this extension comes at a price, that these equations require as an input the knowledge of the partial wave amplitudes for complex values of the energy. It would be interesting to explore their possible phenomenological use.

It would be highly desirable to apply these results to concrete S-matrix bootstrap problems. 
A first possibility is to integrate our formalism into the primal dispersive approach introduced in~\cite{Bhat:2023puy,deRham:2025vaq}.
Lastly, in the dual approach, one expects that dispersion relations valid at all energies will suffer from numerical instabilities, due to the need for analytic continuation when relating partial waves on the imaginary axis to the unitary cut. We conjecture that one could mitigate this issue by introducing a suitable regulator, as proposed in~\cite{He:2021eqn}. We leave these numerical explorations to future works.

\section*{Acknowledgments}

We thank A.~Sinha, K. Haring, A. Homrich, S. Rychkov, and P.~Tourkine for useful discussions.
JEM and AZ are supported by the European Research Council, grant agreement n. 101039756. A.G.
is supported by a Royal Society University Research Fellowship, URF/R1/241371. 

\appendix

\section{Divertissement: Two-channel symmetric dispersion relations}
In this appendix, we describe how to write crossing-symmetric dispersion relations in the case of a two-channel symmetric amplitude $M(s_1,s_2)=M(s_2,s_1).$
The crossing-symmetric variables in this case are defined by 
\be
    x(s_1,s_2) = s_1+s_2, \qquad y(s_1,s_2) = s_1 s_2,
\ee
or via the unordered solutions to the polynomial
\be
    P(\sigma)=(\sigma-\sigma_1)(\sigma-\sigma_2)=\sigma^2 - x \sigma + y = 0.
\ee
The steps of derivation follows analogously to the three channel case on a single leaf $y=f_\alpha(x)$.
To do the pull-back, we first need to solve the equation
\be
    \sigma (x-\sigma)=f_\alpha(x)
    \label{eq:pull_back}
\ee whose solutions are given by $x^{(i)}(\sigma)$ corresponding to $N$ possible solutions. 

We will focus on the open-string amplitude, whose asymptotic behavior needs no subtraction. The un-subtracted dispersion integral is given by
\begin{equation}
   \mathcal{M}_\alpha(x) = \sum_{i=1}^N \frac{1}{\pi} \int_{\mathcal{I}_i} dx' \, \frac{1}{x' - x} \, \mathcal{A}^i_\alpha(x')\, .
\end{equation}
Its pull-back is then given by, with $\tau^{(i)}(\sigma)=x^{(i)}(\sigma)-\sigma$,
\begin{equation}
   M(s_1,s_2)= \sum_{i=1}^N  \frac{1}{\pi} \int_{8/3}^{\infty} d\sigma \, \frac{d x^{(i)}(\sigma)}{d\sigma} \, A(\sigma,\tau^{(i)}(\sigma)) \, \frac{1}{x^{(i)}(\sigma) - s_1 - s_2} 
   \label{eq:two_channel_csdr}
\end{equation}
where
\be
\frac{dx^{(i)}(\sigma)}{d\sigma} = \frac{2\sigma -x}{\sigma-df/dx}\bigg|_{x=x^{(i)}} = \frac{\sigma-\tau^{(i)}(\sigma)}{\sigma-df/dx|_{x=x^{(i)}}} \, .
\ee
The main difference to the three-channel case is that all solutions of $\tau^{(i)}$ are distinct, and there is no notion of ${\pm}$ multiplicity, as there is no $t \leftrightarrow u$ symmetry any more.

For the linear foliation $y=a(x-x_0)$, the generic relation \reef{eq:two_channel_csdr} on a single leaf reduces to
\begin{equation}
   M(s_1,s_2) = \frac{1}{\pi} \int_{8/3}^{\infty} d\sigma \,  A(\sigma,\tau(\sigma)) \left( \frac{1}{\sigma - s_1} + \frac{1}{\sigma - s_2} - \frac{1}{\sigma - a} \right)
   \label{eq:two_channel_linear}
\end{equation}
where 
$$\tau(\sigma)=\frac{a(\sigma-x_0)}{\sigma-a}, \quad x(\sigma)=\frac{\sigma^2-ax_0}{\sigma-a} \, .$$
Note that the un-subtracted kernel after replacing $a=y/(x-x_0)$ becomes
$$
\frac{1}{\sigma - s_1} + \frac{1}{\sigma - s_2} - \frac{s_1+s_2-x_0}{\sigma(s_1+s_2-x_0) - s_1 s_2} \, .
$$

\paragraph{Local CSDR as a limiting case.} As before, we consider the linear foliation with $y=c$ leaves for constant $c$. Taking the double limit $a \to 0$ and $-ax_0 \to c$ gives
$$\tau(\sigma)=\frac{c}{\sigma},\quad x(\sigma)=\frac{c}{\sigma}+\sigma.$$
The kernel~\reef{eq:two_channel_linear} in this limit becomes 
$$\frac{1}{\sigma - s_1} + \frac{1}{\sigma - s_2} - \frac{1}{\sigma},$$
which exactly matches the equation (3) presented in \cite{Saha:2024qpt}, after employing the useful identity
\be
\frac{1}{x(\sigma)-s_1-s_2} = \frac{\sigma-a}{\sigma^2 - (s_1{+}s_2) \sigma + s_1 s_2} \, .
\ee

\subsection{Open string amplitude and a new formula for $\pi$}
The CSDRs on a generic foliation provide new representations for the function 
\be
    M^{\Gamma\Gamma}(s_1,s_2)=\frac{\Gamma \left(\alpha -s_1\right) \Gamma \left(\alpha -s_2\right)}{\Gamma \left(\beta -s_1-s_2\right)} \quad , \quad \alpha,\beta \in \mathbb{R}.
\ee
The absorptive part of this function is given by
\be
    A^{\Gamma\Gamma}(s_1,s_2) = - \sum_{n=0}^\infty \frac{1}{n!} \,\, (1-\alpha +s_2 )_{n+2 \alpha -\beta } \,\, \pi \delta(s_1-n-\alpha)
\ee
where $(x)_n \equiv \Gamma(x+n)/\Gamma(n)$ is the Pochhammer symbol, and when $2\alpha-\beta$ is an integer. 
The open string amplitude corresponds to the case $\alpha=0$, $\beta=1$. 

Employing the pulled-back two-channel CSDRs~\reef{eq:two_channel_csdr}, we get the following representation
\begin{equation}
  M^{\Gamma\Gamma}(s_1,s_2) = \sum_{i=1}^N \sum_{n=0}^{\infty} \, \frac{1}{{n!}}
  {\left(1-\alpha+\tau^{(i)}(\sigma) \right)_{n+2 \alpha -\beta }} \, \times \frac{d x^{(i)}(\sigma)}{d\sigma} \frac{1}{x^{(i)}(\sigma) -s_1-s_2}{\Bigg|}_{\sigma=n+\alpha}
  \label{eq:veneziano_reps}
\end{equation}
provided that the sum over $n$ converges. We remind the reader again that $x^{(i)}(\sigma)$ are the solutions of the equation~\reef{eq:pull_back} where $y=f_{a}(x)$ is a generic foliation.

\begin{figure}[t!] 
\centering 
	\includegraphics[width=0.7\textwidth]{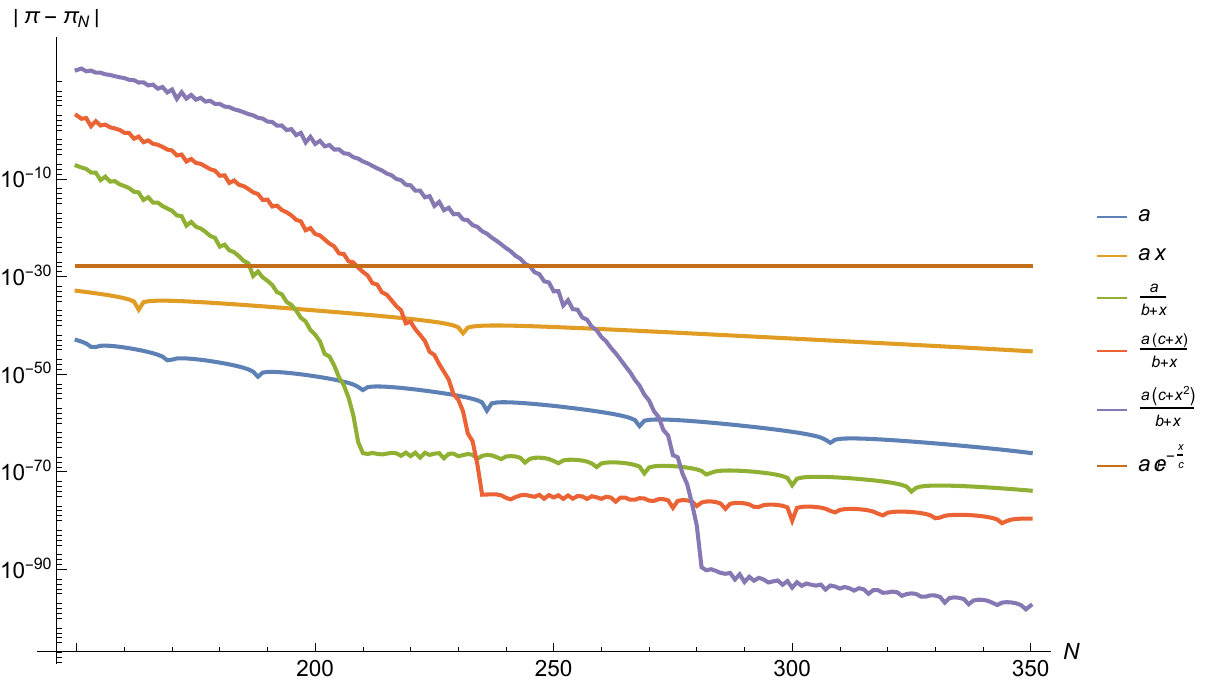} 
	\caption{A comparison of convergence rates of the sum~\reef{eq:new_pi} for various choices of foliations. We fix $b=-201,~c=10, ~\lambda=551/10$ and show the error with respect to to $\pi$ as a function of the truncated $n$-sum.}
\label{pi_N}
\end{figure}

A generic formula for $\pi$ can be found by evaluating the open string amplitude at a kinematic point, by setting $\alpha=0, \, \beta=1, \, s_1=s_2=-\frac{1}{2}$ in~\reef{eq:veneziano_reps}. However, choosing $\alpha=0$ may spoil the convergence of the above sum. For improving the convergence, it is often helpful to use the shift symmetry of this amplitude
$$
    s_1 \to \lambda+s_1, \quad
    s_2 \to \lambda+s_2, \quad
    \alpha \to \alpha+\lambda, \quad
    \beta \to \beta +2 \lambda \, ,
$$
and we finally get 
\be 
    \pi = \sum_{i=1}^N \sum_{n=0}^{\infty} \, \frac{1}{{n!}}{\left(1-\lambda+\tau^{(i)}(\sigma) \right)_{n-1}} \, \times \frac{d x^{(i)}(\sigma)}{d\sigma} \frac{1}{x^{(i)}(\sigma)+1-2\lambda}{\Bigg|}_{\sigma=n+\lambda} \, .
    \label{eq:new_pi}
\ee
For illustration purposes, we considered the following choices of foliations:
$$f_a(x) \in \left\{ a,~ a ~x,~\frac{a}{x+b},~  a~\frac{x+c}{x+b}, ~a~ \frac{x^2+c}{x+b},~  a ~\exp{(-x/c)} \right\}.$$ 
The results are shown in Figure \eqref{pi_N}. The sums generally converge quite fast, reaching up to $70-100$ decimal places accuracy using $<350$ terms in the $n$-sum. 

\paragraph{A worked-out example.} Let us provide an explicit example to show how the general formula works. For the hyperbolic foliation $y=\frac{a}{x+b}$, we get 
\be
    x^{(i)}(\sigma) = \frac{\sigma-b}{2} + \frac{(-)^i}{2} \sqrt{(b+\sigma)^2+\frac{4 a}{\sigma}} 
    \quad \text{and} \quad
    \tau^{(i)}(\sigma)=x^{(i)}(\sigma)-\sigma \, .
\ee 
Then, the un-subtracted kernels are given by
\be
    K^{(i)}(\sigma) = \frac{\sigma^2 \left( b+\sigma+C^{(i)}_\sigma(s_1,s_2) \right) - 2 s_1 s_2 \left(b+s_1+s_2\right)}{\sigma ^2 C^{(i)}_\sigma(s_1,s_2) \left( C^{(i)}_\sigma(s_1,s_2)-b+\sigma -2 s_1-2 s_2 \right) }
\ee
where we defined $C^{(i)}_\sigma(s_1,s_2) \equiv (-)^i\sqrt{\frac{4}{\sigma}s_1 s_2 \left(b+s_1+s_2\right)+(b+\sigma )^2}$.

The hyperbolic foliation, provides a two-parameter representation for the function via $a$ and $b$. 
After using the shift symmetry by $\lambda$, we finally arrive at the representation
\be
M^{\Gamma\Gamma}(s_1,s_2)= - \sum_{i=1}^2 \sum_{n=0}^\infty 
\frac{K^{(i)}(n{+}\alpha)}{n!} \left( 1+\frac{1}{2} C_{n+\alpha+\lambda}(s_1{+}\lambda,s_2{+}\lambda)-3 \alpha -b-3 \lambda-n \right)_{n+2 \alpha -\beta }
\ee
A two-parameter formula for $\pi$ can be obtained once again at $s_1,s_2=-\frac{1}{2}$:
\be 
\pi=-\sum_{n=0}^\infty\left[\frac{\left(1+\frac{\mathcal{C}_n-b-3 \lambda-n}{2} \right)_{n-1}}{2 n!}\left(\frac{\frac{(b+\lambda +n) (b+3 (\lambda +n))}{\mathcal{C}_n (\lambda +n)}-\frac{\mathcal{C}_n}{\lambda +n}+2}{\mathcal{C}_n+n+2-b-3 \lambda}\right)+(\mathcal{C}_n\to-\mathcal{C}_n)\right]
\label{app:pi_sum}
\ee
with $\mathcal{C}_n=\sqrt{(b+\lambda +n)^2+\frac{(1-2 \lambda )^2 (b+2 \lambda -1)}{\lambda +n}}$.

A few words on the convergence of the above sum: For example, choices of $\lambda=551/10$, $b=-200$ with truncation cut off of the $n$-sum at $n_\text{max}=200$, gives $44$ decimal place accuracy to $\pi$. For finite truncation cut off $n_\text{max}$ and positive $b$, the first term in the summand of~\ref{app:pi_sum} contributes the most, e.g. for $\{\lambda,b,n_\text{max}\}=\{551/10,200,200\}$ it reaches $53$ decimal place accuracy to $\pi$.
For negative $b$ both terms are important.

\subsection{Closed string amplitude and multiple zeta function}
To find a crossing-symmetric representation of the closed string amplitude,
\be
M^{\Gamma\Gamma\Gamma}(s,t,u) = \frac{\Gamma (-s) \Gamma (-t) \Gamma (-u)}{\Gamma (s+1) \Gamma (t+1) \Gamma (u+1)} \, ,
\ee
one can use the un-subtracted three-channel CSDR with the $y=c$ foliation, yielding
\be 
M^{\Gamma\Gamma\Gamma}{+}\frac{1}{s t u}=\sum_{n=1}^{\infty} \left(\frac{1}{s-n}+\frac{1}{t-n}+\frac{1}{u-n}+\frac{1}{n}\right) \frac{1}{(n!)^2} \left(1{-}\frac{n}{2}{+}\frac{\sqrt{n^4-4 n s t u}}{2 n}\right)_{n-1}^2
\ee
This expression agrees with \cite{Saha:2024qpt, Bhat:2024agd}, where they used local CSDR to obtain the same expression. Upon using a shift symmetry of the string amplitude, it is possible to get a one-parameter family of representations \cite{Bhat:2024agd,Math}.

Alternatively, using the un-subtracted MRW foliation $y=ax+\beta$, we find
\be
M^{\Gamma\Gamma\Gamma}{+}\frac{1}{s t u}=\sum_{n=1}^{\infty} \left(\frac{1}{n{-}s}{+}\frac{1}{n{-}t}{+}\frac{1}{n{+}s{+}t}{-}\frac{s^2{+}st{+}t^2}{\beta + n \left(s^2{+}st{+}t^2\right)-st(s{+}t)}\right) \frac{\left(\tau(n,\beta)+1\right)_{n-1}^2}{(n!)^2}
\ee
and, with the definitions $x=-s t-tu-us$ and $y=-stu$,
$$\tau(n,\beta) \equiv \frac{1}{2} \frac{1}{\beta +n x-y} \sqrt{(\beta +n x-y) \left(n^3 x+3 n^2 (y-\beta )+4 \beta  x\right)} - \frac{n}{2} \, .$$

These representations both give convergent answers, due to the good large-$n$ behavior of Pochhammer symbols.
We next use them to find parametric representations of the multiple zeta function by expanding on both sides in powers of $s$ and $t$.

\paragraph{Multi-zeta values.}
A linear sum of multiple zeta functions dubbed as $Z(r,q)$ appears in the low energy expansion of the amplitude in terms of $s,t$ as follows (see Equation (4.15) of  \cite{Green:2019tpt} for the exact definition)
\be 
M^{\Gamma\Gamma\Gamma}(s,t,u)+\frac{1}{s t u}=\sum_{i,j=0}^\infty Z(i+3,j) \, s^i t^j\,.
\ee
Then, doing the low energy expansion $M^{\Gamma\Gamma\Gamma}(s,t)=\sum_{l,q=0}^\infty \mathcal{W}_{l,q} x(s,t)^l y(s,t)^q$ in terms of $x,y$ and using the un-subtracted dispersion relation for $y=ax+\beta$, we have 
\be 
\begin{split}
\mathcal{W}_{l,q}^{\Gamma\Gamma\Gamma}=&\sum _{n=1}^{\infty } \sum _{m=0}^q \sum _{j=0}^n \frac{4^j (-1)^{q-m+1}  \beta ^{j-m} P(j,n) n^{-3 j+l-m+3 q-3} }{m! {(n-1)!}^2 (q-m)! \left(\beta +n^3\right)^{l+q-m+1}}\\
&\times  \left(3 \beta  q \, \,  _2\tilde{F}_1\left(1-q,\alpha_l;\beta_j+1;-\frac{\beta }{n^3}\right)+\left(\beta -2 n^3\right) \, _2\tilde{F}_1\left(-q,\alpha_j;\beta_j;-\frac{\beta }{n^3}\right)\right)
\end{split}
\label{app:wlq_sum}
\ee
where $_2\tilde{F}_1$ denotes the regularized Gauss hypergeometric function, and
$$
\alpha_j = j{-}l{+}m{-}q{+}1, \quad 
\beta_j=j{-}q{+}1, \quad
P(j,n)=\frac{\partial ^j}{\partial x^j}\left(\frac{1}{2} n \left(\sqrt{x}-1\right)+1\right)_{n-1}^2\Big|_{x=1}.
$$
Note that the the coefficient of $s^i t^j$ can be extracted from $x(s,t)^l y(s,t)^q=\sum_{i,j=0}^\infty c_{i,j}^{l,q} s^i t^j$ with
\be 
c_{i,j}^{l,q}=\sum_{k=0}^{l} (-1)^q \binom{l}{k} \binom{k}{i-k-q} \binom{q}{-i-j+2 l+3 q} \, .
\label{app:cij_sum}
\ee
Combining~\reef{app:wlq_sum} and~\reef{app:wlq_sum} allows us find the expression for $Z(i+3,j)$. 

Note also that some values of the Riemann zeta function $\zeta$ can be obtained as
\be
\zeta(3+2l)=\mathcal{W}_{l,0}^{\Gamma\Gamma\Gamma} \, .
\ee

\section{Null constraints explicitly}
\label{compcond}

Equation \eqref{eq:pulled_disp_on_non_linear_leaf} describes a one-parameter family of dispersion relations on each leaf $(x,f_\alpha(x))$ covering $\mathbb{C}^2_{(x,y)}$. Stacking them together introduces isolated singular points in the foliation, on which the derivatives with respect to various directions become ill-defined, as described in Section~\ref{nullc}. Such non-analytic behavior in the kinematic plane is spurious since it does not correspond to a physical process. In this section, we explicitly derive the compatibility conditions arising from enforcing analyticity on the singular points for various foliations.

Consider the partial wave expansion in~\reef{eq:pw_expansion} and define $\xi = \cos^2\theta$. To facilitate the expansion around special kinematical points, we can express the Legendre polynomials in terms of a linear combination over simple monomials by noting that 
\be 
P_\ell(\sqrt{\xi})=\sum_{j=0}^{\ell/2}(\xi-\xi_0)^j \frac{P_{j,\ell}}{j!} \, ,
\label{eq:PW_trick}
\ee
where $P_{j,\ell}\equiv\partial^{j}_W P_\ell(\sqrt{W})|_{W=\xi_0}$ for any real $\xi_0$.

Let us explore concrete examples to understand how the constraints arise.

\paragraph{Linear foliation.} Let us consider the linear leaves for constant $x_0$, namely $a(x,y)= y/(x-x_0)$ is a variable of external kinematics.
The scattering angle squared can be recast as
\be
\xi-\xi_0=\frac{4 a \left(\sigma^2-x_0\right)}{\sigma^2 (\sigma-a)} \xi_0 \, .
\ee 
We choose $\xi_0=\frac{\sigma^2}{\left(\sigma-8/3\right)^2}$ to simplify the dependence of the angle on $a$. Plugging the angle into the partial wave expansion~\reef{eq:pw_expansion} and employing \reef{eq:PW_trick} gives 
\be
A(\sigma,s_2^+(\sigma))
= \sum^\infty_{\ell=0} \sum _{j=0}^{\ell/2} \text{Im} f_\ell(\sigma) \left(\frac{4 a \left(\sigma^2-x_0\right)}{\sigma^2 (\sigma-a)}\right)^j (\xi_0)^j \, \frac{P_{j,\ell}}{j!}
\label{app:singular_point}
\ee
After replacing $a \to y/(x-x_0)$, it is obvious that expanding \eqref{app:singular_point} around $x=x_0$ and $y=0$ gives rise to negative powers of $(x-x_0)$. For example:
\be
\reef{app:singular_point} \supset \frac{y^2}{x-x_0} \int_{8/3}^{\infty} \!\! d\sigma \sum^\infty_{\ell=0} \text{Im}f_\ell(\sigma) \, \frac{4 \xi_0 \left(4 \xi_0 P_{2,\ell}-P_{1,\ell}\right)}{\pi  \sigma^5}
\label{eq:null1}
\ee
Such spurious powers to vanish when their coefficients vanish. Imposition of the vanishing of higher-negative powers of $(x-x_0)$ results in generating the null constraints, or locality constraints of Auberson-Khuri CSDR. 

\paragraph{Quadratic foliation.}
A slightly richer example is the foliation $y=f(x)=a x (x-x_0)$, which has two singular points $x=0$ and $x=x_0$, and two solutions for $x^{(i)}$ given by. Plugging the angle in the partial wave expansion, then replacing $a \to y/(x(x-x_0))$ and collecting the negative powers of either $x$ or $(x-x_0)$ produces the null constraints.

For instance, around $(x,y)=(0,0)$, one spurious term is given by
\be
\frac{x}{y} \times \int_{8/3}^{\infty}  d\sigma  \sum^\infty_{\ell=0} \sum _{i=0}^{\ell/2} \text{Im}[f_\ell(\sigma)] 
\frac{{\xi_0}^i P_{i,\ell}}{\pi  (2)_{i-1}}
\left({
\scriptstyle
\frac{\left(\frac{4}{\sigma^2}-4\right)^i \left(i \left(3 \sigma^2-1\right)+6 \sigma^4-7 \sigma^2+1\right)}{\sigma^2-1}-\frac{(-1)^i 2^{2 i+1} \left(3 \sigma^2-2\right) \left(\frac{7}{3}-2 \sigma^2\right){}_{i-1}}{\left(\frac{4}{3}-2 \sigma^2\right){}_{i-1}} 
}\right) \, .
\label{eq:null2}
\ee
\paragraph{Local CSDR.}
An interesting example is the case of constant $y=c$. We showed that every point with a finite distance to the origin is regular in this foliation in Section~\ref{nullc}, however, the parallel leaves intersect at the infinity, giving rise to yet another singularity. We now demonstrate that an expansion around this point naturally leads to null constraints.

Note first that the argument of Legendre polynomials in this case simplify to 
\be 
\xi-\xi_0=\frac{4 c }{\sigma^3} \xi_0 \, .
\ee

Then, for instance, we know that a term of the order $y^n/x$ should not be present in an around $y\to\infty$ while $x$ is fixed, thanks to the Froissart bound, hence the coefficient in front should vanish. The term at this order turns out to be
\be
\frac{y^2}{x} \times \int_{8/3}^{\infty}  d\sigma\sum_{\ell} \text{Im}[f_\ell(\sigma)]\frac{4 \xi_0 \left(4 \xi_0 P_{2,\ell}-P_{1,\ell}\right)}{\pi  \sigma^5} \, .
\label{eq:constant_nc}
\ee
Remarkably, it gives rise to the the same sum rule as~\reef{eq:null1}, as we derived above. 

This resolves a previously puzzling aspect: the local CSDR derived in~\cite{song} must still be supplemented with null constraints, even though they do not emerge directly from the locality property of representation itself. Indeed, the representation appearing in~\cite{song} is fully crossing-symmetric and exhibits a well-defined power expansion in $x$ and $y$, suggesting that the locality is manifest. Thus, when presented with the representation alone, there is no immediate indication of any inconsistency. The necessity for these constraints was not apparent from the representation itself. However, by carefully considering the geometric structure underlying the foliation, the origin of these constraints becomes apparent.

\section{Further details on analyticity domains of linear foliations}
\label{sec:further_det_analyticity}

To determine the allowed region in the \((a, x_0)\)-plane, we fix $x_0$ and scan over $s$ in the three domains: $4 < s < 16$, $16 < s < 32$, and $s > 32$, identifying the values of $a$ for which the dispersive integration remains inside the analyticity domain.

The excluded values of $a$ for each $x_0$ define colored regions in Figure~\ref{ax0regions} (left), while the remaining white region corresponds to the admissible choices of $(a, x_0)$. The optimal value is found at $x_0 = 787.6$, where the allowed range is $-44.4 < a < 8/3$. For the Auberson-Khuri choice $x_0 = 0$, the range is narrower: $-29.5 < a < 8/3$—see also \cite{Chowdhury:2022obp}. These bounds translate into maximum energies of validity for the Roy equations: $s = 125.31$ and $s = 89.9$, respectively.

\paragraph{Linear foliation.}

Using Figure~\ref{ellipses_Martin}, we can now translate the $t$-analyticity domain into constraints on 
the foliation parameters used to derive CSDRs. For instance, the Landau curve $t = \frac{4s}{s - 16}$ maps to the straight line $y = \frac{8}{3}x$. Consequently, any dispersive integration along a line $y = a(x - x_0)$ with $a \geq \frac{8}{3}$ will extend beyond the analyticity domain.

An arbitrary choice of \((a, x_0)\) is not consistent with the analyticity domain \(\mathcal{D}^\prime\) of~\eqref{larger_ellipses}, so let us identify the allowed regions in \((a, x_0)\) space. These domains are defined through Lehmann–Martin ellipses \eqref{eq:elip} whose right extremity $r'(s)$ were given in \eqref{extremities}. 
Ensuring that $s_2^+(s_1)$ in~\reef{eq:s2_plusminus} stays inside the ellipses eventually leads to a complex domain for $a$ at a fixed $x_0$. Since are interested in the real values of $a$, we keep $\im a=0$. The key steps to find the allowed regions are:
\begin{itemize}
    \item At a fixed \(x_0\), find the maximum and minimum of $a$ allowed by the ellipses, scanning over the $s$ in three regions: \(4 < s < 16\), \(16 < s < 32\), and \(s > 32\). 
    \item Identify excluded regions (yellow, green, cyan in left panel of Figure~\ref{ax0regions}) and allowed regions (white) in the \((a, x_0)\)-plane, namely for each $x_0$ find allowed-disallowed values of $a$ and plot them as a function of $x_0$.
\end{itemize}
The figure shows a global minimum for optimal value \(x_0 = 787.6\) with the allowed range \(-44.4 < a < 8/3\). For the Auberson-Khuri choice, \(x_0 = 0\), the allowed range is \(-29.5 < a < 8/3\)-----see also the derivation in~\cite{Chowdhury:2022obp}. The corresponding upper bounds for the validity domain of Roy equations are respectively \(s = 125.31\) and \(s = 89.9\). 

\paragraph{Local CSDR.}
For the foliation of constant lines, \(y = c\), the arguments of absorptive part are $s_1$ and $s_2^+(s_1)$ given by~\reef{eq:lcsdr_s2}.
To ensure analyticity, we must demand that $s_2^+(s_1)$ lies within the Lehmann-Martin ellipses \eqref{eq:elip}. This leads to a complex domain for $c$. 
In our foliation, we are interested in the real values of $c$, so we keep $\im c=0$. We repeat the same exercise as above:     
    \begin{itemize}
    \item Find the maximum and minimum of \(c\) allowed by the ellipses scanning over $s$ in three ranges: \(4 < s < 16\), \(16 < s < 32\), and \(s > 32\). The second ellipse $16<s<32$ gives the most stringent bounds
    \item The allowed values of $c$ lie in the interval
    $$
    -\frac{198704}{27} < c < \frac{170368}{27}.
    $$
    \end{itemize}
Then, the validity domain of the Roy equation in this case turns out to be \(-28<s<32.2103\).

\section{Further details on Roy-like equations for all energies}
\label{details_roy_all}

Let us first briefly show that why Roy-like equations we obtained in Section~\ref{sec:homogeneous_roy} via the homogeneous foliation \eqref{eq:homogeneous_foliation} works for all physical energies and angles. Remember that the arguments of $\mathcal{M}(x,y)$ should obey the inequality 
$$-\frac{2}{3\sqrt{3}}<\frac{y(s_1,s_2)}{x(s_1,s_2)^{3/2}}<0$$
to avoid the forbidden regions in $(x,y)$ plane. If we replace now $t=-\frac{1}{2} (s-4) (1-z)$, then the above inequality becomes
\be
-\frac{2}{3\sqrt{3}}<\frac{18 (s-4)^2 (3 s-4) z^2-2 (3 s-4)^3}{3 \sqrt{3} \left(3 (s-4)^2 z^2+(4-3 s)^2\right)^{3/2}}<0 \, ,
\label{eq:alpharange}
\ee
which is always respected for physical kinematics, i.e. $s\geq 4$ and $-1\leq z\leq 1$. In other words, there is no upper bound on $s$ for the validity range of Roy-like equations.

Then, let us move on deriving the Roy-like equations for all energies.

The unique solution to $\sigma^3 - \sigma x + \alpha x^{3/2} = 0$  for $\alpha <0$ is given by

\begin{equation}
    x_\alpha(\sigma) = \frac{\sigma ^2 \left(2 \sqrt[3]{2} \left(6 \alpha ^2-1\right)+2 \sqrt[3]{c_{\alpha }}-2^{2/3} c_{\alpha }^{2/3}\right)}{6 \alpha ^2 \sqrt[3]{c_{\alpha }}}\,,
\end{equation}
where the constant $c_{\alpha }= -3 \left(\alpha  \left(\sqrt{81 \alpha ^2-12}+9 \alpha \right)-6\right) \alpha ^2-2$ depends on $\alpha$. 
We also recall that $\tau_\alpha(\sigma) = -\tfrac{\sigma}{2} + \tfrac{1}{2} \sqrt{4x_\alpha(\sigma) - 3\sigma^2}$.
The first step is to expand the integrand of \eqref{eq:nonanalytic_csdr} in partial waves, which yields
\begin{eqnarray}
M(s_1,s_2)=M(0,0)+\frac{1}{\pi}\int_{\tfrac{8}{3}}^\infty d\sigma \sum_{\ell=0}^\infty 16\pi(2\ell+1) \, \im \left[ f_\ell(\sigma) \right] P_\ell(\tfrac{\sigma+2\tau_\alpha(\sigma)}{\sigma-8/3})\frac{2s_1^2}{\sigma^3-s_1^2\sigma}\nn\\
+\frac{1}{\pi}\int_{0}^\infty d\sigma \sum_{\ell=0}^\infty 16\pi(2\ell+1) \, \im \left[ f_\ell(i\sigma)P_\ell(\tfrac{i\sigma+2\tau_\alpha(i\sigma)}{i\sigma-8/3}) \right]\frac{2s_1^2}{\sigma^3+s_1^2\sigma},
\end{eqnarray}
where we used the fact that for $\alpha<0$, $\tau_\alpha(-i\sigma)=\tau_\alpha(i\sigma)^*$, and real analyticity $f_\ell(-i\sigma)=f_\ell(i\sigma)^*$.
Partial wave projection is given by
\be
    f_\ell(s_1)=\frac{1}{32\pi}\int_{-1}^1 dz P_\ell(z)M(s_1,s_2) \quad \text{where} \, z=\tfrac{s_1+2s_2}{s_1-8/3}\, .
\ee
Note that, for any external $s_1$, we integrate over the full range of $z$ by adjusting $\alpha = \frac{y(s_1,s_2)}{x(s_1,s_2)^{3/2}} < 0$ accordingly.
Then we have the final form of Roy-like equations as follows
\begin{eqnarray}
f_\ell(s_1)=\frac{M(0,0)\delta_{\ell,0}}{16\pi}+\int_{\tfrac{8}{3}}^\infty d\sigma \sum_{\ell'=0}^\infty \im f_{\ell'}(\sigma)K^1_{\ell',\ell}(\sigma,s_1)\nn\\
+\int_0^\infty d\sigma \sum_{\ell'=0}^\infty (\im f_{\ell'}(i\sigma) K^2_{\ell',\ell}(\sigma,s_1)+\re f_{\ell'}(i\sigma)K^3_{\ell',\ell}(\sigma,s_1)),
\end{eqnarray}
where the kernels are given by
\be
    K^i_{\ell',\ell}(\sigma,s_1)=\frac{2\ell'+1}{2 \pi} \int_{-1}^1 dz \, P_\ell(z) \, k^i_{\ell'}(s_1,z,\sigma)
\ee
and
\begin{eqnarray}
    k^1_{\ell'}(s_1,z,\sigma)=P_{\ell'}\left(\frac{\sigma+2\tau_\alpha(\sigma)}{\sigma-8/3}\right)\frac{2s^2_1}{\sigma^3 - s_1^2 \sigma}\nn\\
    k^2_{\ell'}(s_1,z,\sigma)=\re P_{\ell'}\left(\frac{i\sigma+2\tau_\alpha(i \sigma)}{i\sigma-8/3}\right)\frac{2s^2_1}{\sigma^3+s_1^2 \sigma}\nn\\
    k^3_{\ell'}(s_1,z,\sigma)=\im P_{\ell'}\left(\frac{i\sigma+2\tau_\alpha(i\sigma)}{i\sigma-8/3}\right)\frac{2s^2_1}{\sigma^3+s_1^2 \sigma}\nn
\end{eqnarray}
The $z$–dependence of the integrands comes through $\alpha= \frac{y(s_1,s_2)}{x(s_1,s_2)^{3/2}}$ with $z=\tfrac{s_1+2s_2}{s_1-8/3}$---see also \eqref{eq:alpharange}.

\subsection{Convergence of partial wave expansion along imaginary axis}

Let us fix the ratio $s_2/s_1=k=$ const. first and we obtain $x(\sigma, k\sigma) = \sigma^2(1+k+k^2)$. This expression for $x$ becomes useful when plugged in $\sigma^3 - \sigma x + \alpha x^{3/2} = 0$, because it allows to express $\alpha$ in terms of the ratio of Mandelstam variables
\be
\sigma^3(k+k^2) = \alpha \sigma^3 (1+k+k^2)^{3/2} \implies \alpha = \frac{k+k^2}{(1+k+k^2)^{3/2}} \, .
\ee
The fundamental domain is fully spanned by $k \in [-1/2,1]$ -- see Figure~\ref{fig_geometry}. Note that the restriction \reef{eq:roy_infinity} implies that
\be
-1/2 \leq k \leq 0 \, ,
\label{eq:k_range}
\ee
and the map between $k$ and $\alpha$ is one-to-one in this interval. 

In order to write Roy-like equations for this foliation, we will need to expand $M(\sigma,k\sigma)$ in partial waves along imaginary axis in $\sigma$. The scattering angle on the leaf is given by
\be
z = \frac{s_2-s_3}{s_1 - 8/3} = \frac{(2k+1)\,\sigma}{\sigma-8/3}
\ee
The closest singularity in the complex $z$ plane is given by $z_\text{small}=1+\frac{8}{\sigma-8/3}$, and the small Martin-Lehmann ellipse whose foci are at $z=\pm 1$ passes through this point. 

From now on, we assume $\sigma = i \bar{\sigma} \in i \mathbb{R}$, then the interior of the ellipse is given by
\be
\frac{(\re z)^2}{a^2} + \frac{(\im z)^2}{a^2-1} \leq 1 \quad \text{where} \, a = \frac{4+\sqrt{\bar{\sigma}^2+16/9}}{\sqrt{\bar{\sigma}^2+64/9}} \, .
\ee
After plugging in the scattering angle along the imaginary line, we get to the inequality
\be
\frac{3 \bar{\sigma} ^2 \left(3 \sqrt{9 \bar{\sigma} ^2+16}+20\right)}{\left(\sqrt{9 \bar{\sigma} ^2+16}+4\right) \left(\sqrt{9 \bar{\sigma} ^2+16}+12\right)^2} \leq \frac{1}{(1+2k)^2} \, .
\label{eq:roy_lehmann_ineq}
\ee
Left-hand side is a monotonic function of $\bar{\sigma} \in [0,\infty]$ and it is bounded by 1, which makes the above inequality always true since right-hand side is greater than 1 when $k$ is in the range \reef{eq:roy_infinity}. Hence, the partial wave expansion converges. Remark also that it stops converging for any $\alpha > 0$.

\bibliographystyle{utphys}
\bibliography{biblio}

\providecommand{\href}[2]{#2}\begingroup\raggedright\begin{thebibliography}{10}

\bibitem{Auberson:1972prg}
G.~Auberson and N.~N. Khuri, ``{Rigorous parametric dispersion representation with three-channel symmetry},'' \href{http://dx.doi.org/10.1103/PhysRevD.6.2953}{{\em Phys. Rev. D} {\bfseries 6} (1972) 2953--2966}.

\bibitem{Mahoux:1974ej}
G.~Mahoux, S.~M. Roy, and G.~Wanders, ``{Physical pion pion partial-wave equations based on three channel crossing symmetry},'' \href{http://dx.doi.org/10.1016/0550-3213(74)90480-5}{{\em Nucl. Phys. B} {\bfseries 70} (1974) 297--316}.

\bibitem{Atkinson:1974ev}
D.~Atkinson and T.~P. Pool, ``{Study of the Roy Equations. 1. Analyticity, Crossing and Threshold Behavior},'' \href{http://dx.doi.org/10.1016/0550-3213(74)90247-8}{{\em Nucl. Phys. B} {\bfseries 81} (1974) 502--516}.

\bibitem{Sinha:2020win}
A.~Sinha and A.~Zahed, ``{Crossing Symmetric Dispersion Relations in Quantum Field Theories},'' \href{http://dx.doi.org/10.1103/PhysRevLett.126.181601}{{\em Phys. Rev. Lett.} {\bfseries 126} no.~18, (2021) 181601}, \href{http://arxiv.org/abs/2012.04877}{{\ttfamily arXiv:2012.04877 [hep-th]}}.

\bibitem{Gopakumar:2021dvg}
R.~Gopakumar, A.~Sinha, and A.~Zahed, ``{Crossing Symmetric Dispersion Relations for Mellin Amplitudes},'' \href{http://dx.doi.org/10.1103/PhysRevLett.126.211602}{{\em Phys. Rev. Lett.} {\bfseries 126} no.~21, (2021) 211602}, \href{http://arxiv.org/abs/2101.09017}{{\ttfamily arXiv:2101.09017 [hep-th]}}.

\bibitem{Saha:2024qpt}
A.~P. Saha and A.~Sinha, ``{Field Theory Expansions of String Theory Amplitudes},'' \href{http://dx.doi.org/10.1103/PhysRevLett.132.221601}{{\em Phys. Rev. Lett.} {\bfseries 132} no.~22, (2024) 221601}, \href{http://arxiv.org/abs/2401.05733}{{\ttfamily arXiv:2401.05733 [hep-th]}}.

\bibitem{Bhat:2024agd}
F.~Bhat, D.~Chowdhury, A.~P. Saha, and A.~Sinha, ``{Bootstrapping string models with entanglement minimization and machine learning},'' \href{http://dx.doi.org/10.1103/PhysRevD.111.066013}{{\em Phys. Rev. D} {\bfseries 111} no.~6, (2025) 066013}, \href{http://arxiv.org/abs/2409.18259}{{\ttfamily arXiv:2409.18259 [hep-th]}}.

\bibitem{Bhat:2025zex}
F.~Bhat, A.~P. Saha, and A.~Sinha, ``{A stringy dispersion relation for field theory},'' \href{http://arxiv.org/abs/2506.03862}{{\ttfamily arXiv:2506.03862 [hep-th]}}.

\bibitem{deRham:2022gfe}
C.~de~Rham, S.~Jaitly, and A.~J. Tolley, ``{Constraints on Regge behavior from IR physics},'' \href{http://dx.doi.org/10.1103/PhysRevD.108.046011}{{\em Phys. Rev. D} {\bfseries 108} no.~4, (2023) 046011}, \href{http://arxiv.org/abs/2212.04975}{{\ttfamily arXiv:2212.04975 [hep-th]}}.

\bibitem{Chang:2025cxc}
C.-H. Chang and J.~Parra-Martinez, ``{Graviton loops and negativity},'' \href{http://dx.doi.org/10.1007/JHEP08(2025)175}{{\em JHEP} {\bfseries 08} (2025) 175}, \href{http://arxiv.org/abs/2501.17949}{{\ttfamily arXiv:2501.17949 [hep-th]}}.

\bibitem{Pasiecznik:2025eqc}
C.~Pasiecznik, ``{Bootstrapping Gravity with Crossing Symmetric Dispersion Relations},'' \href{http://arxiv.org/abs/2506.09884}{{\ttfamily arXiv:2506.09884 [hep-th]}}.

\bibitem{Alday:2022xwz}
L.~F. Alday, T.~Hansen, and J.~A. Silva, ``{AdS Virasoro-Shapiro from single-valued periods},'' \href{http://dx.doi.org/10.1007/JHEP12(2022)010}{{\em JHEP} {\bfseries 12} (2022) 010}, \href{http://arxiv.org/abs/2209.06223}{{\ttfamily arXiv:2209.06223 [hep-th]}}.

\bibitem{Bhat:2023ekh}
F.~Bhat and A.~Zahed, ``{A Celestial route to AdS bulk locality},'' \href{http://dx.doi.org/10.1007/JHEP08(2023)112}{{\em JHEP} {\bfseries 08} (2023) 112}, \href{http://arxiv.org/abs/2304.02003}{{\ttfamily arXiv:2304.02003 [hep-th]}}.

\bibitem{Bissi:2022fmj}
A.~Bissi and A.~Sinha, ``{Positivity, low twist dominance and CSDR for CFTs},'' \href{http://dx.doi.org/10.21468/SciPostPhys.14.4.083}{{\em SciPost Phys.} {\bfseries 14} no.~4, (2023) 083}, \href{http://arxiv.org/abs/2209.03978}{{\ttfamily arXiv:2209.03978 [hep-th]}}.

\bibitem{Chowdhury:2021ynh}
S.~D. Chowdhury, K.~Ghosh, P.~Haldar, P.~Raman, and A.~Sinha, ``{Crossing Symmetric Spinning S-matrix Bootstrap: EFT bounds},'' \href{http://dx.doi.org/10.21468/SciPostPhys.13.3.051}{{\em SciPost Phys.} {\bfseries 13} no.~3, (2022) 051}, \href{http://arxiv.org/abs/2112.11755}{{\ttfamily arXiv:2112.11755 [hep-th]}}.

\bibitem{Zahed:2021fkp}
A.~Zahed, ``{Positivity and geometric function theory constraints on pion scattering},'' \href{http://dx.doi.org/10.1007/JHEP12(2021)036}{{\em JHEP} {\bfseries 12} (2021) 036}, \href{http://arxiv.org/abs/2108.10355}{{\ttfamily arXiv:2108.10355 [hep-th]}}.

\bibitem{Li:2023qzs}
Y.-Z. Li, ``{Effective field theory bootstrap, large-N {\ensuremath{\chi}}PT and holographic QCD},'' \href{http://dx.doi.org/10.1007/JHEP01(2024)072}{{\em JHEP} {\bfseries 01} (2024) 072}, \href{http://arxiv.org/abs/2310.09698}{{\ttfamily arXiv:2310.09698 [hep-th]}}.

\bibitem{Haldar:2021rri}
P.~Haldar, A.~Sinha, and A.~Zahed, ``{Quantum field theory and the Bieberbach conjecture},'' \href{http://dx.doi.org/10.21468/SciPostPhys.11.1.002}{{\em SciPost Phys.} {\bfseries 11} (2021) 002}, \href{http://arxiv.org/abs/2103.12108}{{\ttfamily arXiv:2103.12108 [hep-th]}}.

\bibitem{Raman:2021pkf}
P.~Raman and A.~Sinha, ``{QFT, EFT and GFT},'' \href{http://dx.doi.org/10.1007/JHEP12(2021)203}{{\em JHEP} {\bfseries 12} (2021) 203}, \href{http://arxiv.org/abs/2107.06559}{{\ttfamily arXiv:2107.06559 [hep-th]}}.

\bibitem{Paulos:2017fhb}
M.~F. Paulos, J.~Penedones, J.~Toledo, B.~C. van Rees, and P.~Vieira, ``{The S-matrix bootstrap. Part III: higher dimensional amplitudes},'' \href{http://dx.doi.org/10.1007/JHEP12(2019)040}{{\em JHEP} {\bfseries 12} (2019) 040}, \href{http://arxiv.org/abs/1708.06765}{{\ttfamily arXiv:1708.06765 [hep-th]}}.

\bibitem{Guerrieri:2018uew}
A.~L. Guerrieri, J.~Penedones, and P.~Vieira, ``{Bootstrapping QCD Using Pion Scattering Amplitudes},'' \href{http://dx.doi.org/10.1103/PhysRevLett.122.241604}{{\em Phys. Rev. Lett.} {\bfseries 122} no.~24, (2019) 241604}, \href{http://arxiv.org/abs/1810.12849}{{\ttfamily arXiv:1810.12849 [hep-th]}}.

\bibitem{Guerrieri:2020bto}
A.~Guerrieri, J.~Penedones, and P.~Vieira, ``{S-matrix Bootstrap for Effective Field Theories: Massless Pions},'' \href{http://arxiv.org/abs/2011.02802}{{\ttfamily arXiv:2011.02802 [hep-th]}}.

\bibitem{Hebbar:2020ukp}
A.~Hebbar, D.~Karateev, and J.~Penedones, ``{Spinning S-matrix Bootstrap in 4d},'' \href{http://arxiv.org/abs/2011.11708}{{\ttfamily arXiv:2011.11708 [hep-th]}}.

\bibitem{Guerrieri:2021ivu}
A.~Guerrieri, J.~Penedones, and P.~Vieira, ``{Where Is String Theory in the Space of Scattering Amplitudes?},'' \href{http://dx.doi.org/10.1103/PhysRevLett.127.081601}{{\em Phys. Rev. Lett.} {\bfseries 127} no.~8, (2021) 081601}, \href{http://arxiv.org/abs/2102.02847}{{\ttfamily arXiv:2102.02847 [hep-th]}}.

\bibitem{He:2021eqn}
Y.~He and M.~Kruczenski, ``{S-matrix bootstrap in 3+1 dimensions: regularization and dual convex problem},'' \href{http://dx.doi.org/10.1007/JHEP08(2021)125}{{\em JHEP} {\bfseries 08} (2021) 125}, \href{http://arxiv.org/abs/2103.11484}{{\ttfamily arXiv:2103.11484 [hep-th]}}.

\bibitem{Chen:2022nym}
H.~Chen, A.~L. Fitzpatrick, and D.~Karateev, ``{Nonperturbative bounds on scattering of massive scalar particles in d \ensuremath{\geq} 2},'' \href{http://dx.doi.org/10.1007/JHEP12(2022)092}{{\em JHEP} {\bfseries 12} (2022) 092}, \href{http://arxiv.org/abs/2207.12448}{{\ttfamily arXiv:2207.12448 [hep-th]}}.

\bibitem{EliasMiro:2022xaa}
J.~Elias~Miro, A.~Guerrieri, and M.~A. Gumus, ``{Bridging Positivity and S-matrix Bootstrap Bounds},'' \href{http://arxiv.org/abs/2210.01502}{{\ttfamily arXiv:2210.01502 [hep-th]}}.

\bibitem{Guerrieri:2022sod}
A.~Guerrieri, H.~Murali, J.~Penedones, and P.~Vieira, ``{Where is M-theory in the space of scattering amplitudes?},'' \href{http://dx.doi.org/10.1007/JHEP06(2023)064}{{\em JHEP} {\bfseries 06} (2023) 064}, \href{http://arxiv.org/abs/2212.00151}{{\ttfamily arXiv:2212.00151 [hep-th]}}.

\bibitem{Tourkine:2023xtu}
P.~Tourkine and A.~Zhiboedov, ``{Scattering amplitudes from dispersive iterations of unitarity},'' \href{http://dx.doi.org/10.1007/JHEP11(2023)005}{{\em JHEP} {\bfseries 11} (2023) 005}, \href{http://arxiv.org/abs/2303.08839}{{\ttfamily arXiv:2303.08839 [hep-th]}}.

\bibitem{Acanfora:2023axz}
F.~Acanfora, A.~Guerrieri, K.~H{\"a}ring, and D.~Karateev, ``{Bounds on scattering of neutral Goldstones},'' \href{http://dx.doi.org/10.1007/JHEP03(2024)028}{{\em JHEP} {\bfseries 03} (2024) 028}, \href{http://arxiv.org/abs/2310.06027}{{\ttfamily arXiv:2310.06027 [hep-th]}}.

\bibitem{Gumus:2024lmj}
M.~A. Gumus, D.~Leflot, P.~Tourkine, and A.~Zhiboedov, ``{The S-matrix bootstrap with neural optimizers. Part I. Zero double discontinuity},'' \href{http://dx.doi.org/10.1007/JHEP07(2025)210}{{\em JHEP} {\bfseries 07} (2025) 210}, \href{http://arxiv.org/abs/2412.09610}{{\ttfamily arXiv:2412.09610 [hep-th]}}.

\bibitem{Bhat:2023puy}
F.~Bhat, D.~Chowdhury, A.~Sinha, S.~Tiwari, and A.~Zahed, ``{Bootstrapping high-energy observables},'' \href{http://dx.doi.org/10.1007/JHEP03(2024)157}{{\em JHEP} {\bfseries 03} (2024) 157}, \href{http://arxiv.org/abs/2311.03451}{{\ttfamily arXiv:2311.03451 [hep-th]}}.

\bibitem{Guerrieri:2024jkn}
A.~Guerrieri, K.~H{\"a}ring, and N.~Su, ``{From data to the analytic S-matrix: A Bootstrap fit of the pion scattering amplitude},'' \href{http://arxiv.org/abs/2410.23333}{{\ttfamily arXiv:2410.23333 [hep-th]}}.

\bibitem{deRham:2025vaq}
C.~de~Rham, A.~J. Tolley, Z.-H. Wang, and S.-Y. Zhou, ``{Primal S-matrix bootstrap with dispersion relations},'' \href{http://arxiv.org/abs/2506.22546}{{\ttfamily arXiv:2506.22546 [hep-th]}}.

\bibitem{Correia:2025uvc}
M.~Correia, A.~Georgoudis, and A.~L. Guerrieri, ``{Cross-Section Bootstrap: Unveiling the Froissart Amplitude},'' \href{http://arxiv.org/abs/2506.04313}{{\ttfamily arXiv:2506.04313 [hep-th]}}.

\bibitem{Lopez:1974cq}
C.~Lopez, ``{A Lower Bound to the pi0 pi0 S-Wave Scattering Length},'' \href{http://dx.doi.org/10.1016/0550-3213(75)90287-4}{{\em Nucl. Phys. B} {\bfseries 88} (1975) 358--364}.

\bibitem{Lopez:1976zs}
C.~Lopez and G.~Mennessier, ``{Bounds on the pi0 pi0 Amplitude},'' \href{http://dx.doi.org/10.1016/0550-3213(77)90237-1}{{\em Nucl. Phys. B} {\bfseries 118} (1977) 426--444}.

\bibitem{Guerrieri:2021tak}
A.~Guerrieri and A.~Sever, ``{Rigorous Bounds on the Analytic S Matrix},'' \href{http://dx.doi.org/10.1103/PhysRevLett.127.251601}{{\em Phys. Rev. Lett.} {\bfseries 127} no.~25, (2021) 251601}, \href{http://arxiv.org/abs/2106.10257}{{\ttfamily arXiv:2106.10257 [hep-th]}}.

\bibitem{EliasMiro:2023fqi}
J.~Elias~Miro, A.~Guerrieri, and M.~A. Gumus, ``{Extremal Higgs couplings},'' \href{http://arxiv.org/abs/2311.09283}{{\ttfamily arXiv:2311.09283 [hep-ph]}}.

\bibitem{Guerrieri:2023qbg}
A.~L. Guerrieri, A.~Hebbar, and B.~C. van Rees, ``{Constraining Glueball Couplings},'' \href{http://arxiv.org/abs/2312.00127}{{\ttfamily arXiv:2312.00127 [hep-th]}}.

\bibitem{Auberson:1974in}
G.~Auberson and L.~Epele, ``{A Tool for Extending the Analyticity Domain of Partial Wave Amplitudes and the Validity of Roy-Type Equations},'' \href{http://dx.doi.org/10.1007/BF02820858}{{\em Nuovo Cim. A} {\bfseries 25} (1975) 453}.

\bibitem{Roy:1971tc}
S.~M. Roy, ``{Exact integral equation for pion pion scattering involving only physical region partial waves},'' \href{http://dx.doi.org/10.1016/0370-2693(71)90724-6}{{\em Phys. Lett. B} {\bfseries 36} (1971) 353--356}.

\bibitem{Auberson:1977re}
G.~Auberson and S.~Ciulli, ``{A Set of Integral Equations for Pion Pion Scattering Valid at All Energies},'' \href{http://dx.doi.org/10.1007/BF02896339}{{\em Nuovo Cim. A} {\bfseries 44} (1978) 549}.

\bibitem{Wanders:1996}
G.~Wanders, ``{Sum rules for pi–pi scattering},'' {\em Helv. Phys. Acta} {\bfseries 39} (1966) 228--246.

\bibitem{Martin:1965jj}
A.~Martin, ``{Extension of the axiomatic analyticity domain of scattering amplitudes by unitarity. 1.},'' \href{http://dx.doi.org/10.1007/BF02720568}{{\em Nuovo Cim. A} {\bfseries 42} (1965) 930--953}.

\bibitem{Martin:1966zsy}
A.~Martin, ``{Extension of the axiomatic analyticity domain of scattering amplitudes by unitarity.{\textemdash}II},'' \href{http://dx.doi.org/10.1007/bf02719361}{{\em Nuovo Cim. A} {\bfseries 44} no.~4, (1966) 1219--1244}.

\bibitem{Jin:1964zza}
Y.~S. Jin and A.~Martin, ``{Number of Subtractions in Fixed-Transfer Dispersion Relations},'' \href{http://dx.doi.org/10.1103/PhysRev.135.B1375}{{\em Phys. Rev.} {\bfseries 135} (1964) B1375--B1377}.

\bibitem{song}
C.~Song, ``{Crossing-Symmetric Dispersion Relations without Spurious Singularities},'' \href{http://dx.doi.org/10.1103/PhysRevLett.131.161602}{{\em Phys. Rev. Lett.} {\bfseries 131} no.~16, (2023) 161602}, \href{http://arxiv.org/abs/2305.03669}{{\ttfamily arXiv:2305.03669 [hep-th]}}.

\bibitem{Roskies:1970uj}
R.~Roskies, ``{Crossing restrictions on pi pi partial waves},'' \href{http://dx.doi.org/10.1007/BF02824912}{{\em Nuovo Cim. A} {\bfseries 65} (1970) 467--490}.

\bibitem{Ananthanarayan:2000ht}
B.~Ananthanarayan, G.~Colangelo, J.~Gasser, and H.~Leutwyler, ``{Roy equation analysis of pi pi scattering},'' \href{http://dx.doi.org/10.1016/S0370-1573(01)00009-6}{{\em Phys. Rept.} {\bfseries 353} (2001) 207--279}, \href{http://arxiv.org/abs/hep-ph/0005297}{{\ttfamily arXiv:hep-ph/0005297}}.

\bibitem{Tolley:2020gtv}
A.~J. Tolley, Z.-Y. Wang, and S.-Y. Zhou, ``{New positivity bounds from full crossing symmetry},'' \href{http://arxiv.org/abs/2011.02400}{{\ttfamily arXiv:2011.02400 [hep-th]}}.

\bibitem{Caron-Huot:2020cmc}
S.~Caron-Huot and V.~Van~Duong, ``{Extremal Effective Field Theories},'' \href{http://arxiv.org/abs/2011.02957}{{\ttfamily arXiv:2011.02957 [hep-th]}}.

\bibitem{Albert:2022oes}
J.~Albert and L.~Rastelli, ``{Bootstrapping pions at large N},'' \href{http://dx.doi.org/10.1007/JHEP08(2022)151}{{\em JHEP} {\bfseries 08} (2022) 151}, \href{http://arxiv.org/abs/2203.11950}{{\ttfamily arXiv:2203.11950 [hep-th]}}.

\bibitem{Albert:2023jtd}
J.~Albert and L.~Rastelli, ``{Bootstrapping pions at large N. Part II. Background gauge fields and the chiral anomaly},'' \href{http://dx.doi.org/10.1007/JHEP09(2024)039}{{\em JHEP} {\bfseries 09} (2024) 039}, \href{http://arxiv.org/abs/2307.01246}{{\ttfamily arXiv:2307.01246 [hep-th]}}.

\bibitem{Albert:2023seb}
J.~Albert, J.~Henriksson, L.~Rastelli, and A.~Vichi, ``{Bootstrapping mesons at large N: Regge trajectory from spin-two maximization},'' \href{http://dx.doi.org/10.1007/JHEP09(2024)172}{{\em JHEP} {\bfseries 09} (2024) 172}, \href{http://arxiv.org/abs/2312.15013}{{\ttfamily arXiv:2312.15013 [hep-th]}}.

\bibitem{Albert:2024yap}
J.~Albert, W.~Knop, and L.~Rastelli, ``{Where is tree-level string theory?},'' \href{http://dx.doi.org/10.1007/JHEP02(2025)157}{{\em JHEP} {\bfseries 02} (2025) 157}, \href{http://arxiv.org/abs/2406.12959}{{\ttfamily arXiv:2406.12959 [hep-th]}}.

\bibitem{Bern:2021ppb}
Z.~Bern, D.~Kosmopoulos, and A.~Zhiboedov, ``{Gravitational effective field theory islands, low-spin dominance, and the four-graviton amplitude},'' \href{http://dx.doi.org/10.1088/1751-8121/ac0e51}{{\em J. Phys. A} {\bfseries 54} no.~34, (2021) 344002}, \href{http://arxiv.org/abs/2103.12728}{{\ttfamily arXiv:2103.12728 [hep-th]}}.

\bibitem{Caron-Huot:2022ugt}
S.~Caron-Huot, Y.-Z. Li, J.~Parra-Martinez, and D.~Simmons-Duffin, ``{Causality constraints on corrections to Einstein gravity},'' \href{http://dx.doi.org/10.1007/JHEP05(2023)122}{{\em JHEP} {\bfseries 05} (2023) 122}, \href{http://arxiv.org/abs/2201.06602}{{\ttfamily arXiv:2201.06602 [hep-th]}}.

\bibitem{Fernandez:2022kzi}
C.~Fernandez, A.~Pomarol, F.~Riva, and F.~Sciotti, ``{Cornering large-N$_{c}$ QCD with positivity bounds},'' \href{http://dx.doi.org/10.1007/JHEP06(2023)094}{{\em JHEP} {\bfseries 06} (2023) 094}, \href{http://arxiv.org/abs/2211.12488}{{\ttfamily arXiv:2211.12488 [hep-th]}}.

\bibitem{Ma:2023vgc}
T.~Ma, A.~Pomarol, and F.~Sciotti, ``{Bootstrapping the chiral anomaly at large N$_{c}$},'' \href{http://dx.doi.org/10.1007/JHEP11(2023)176}{{\em JHEP} {\bfseries 11} (2023) 176}, \href{http://arxiv.org/abs/2307.04729}{{\ttfamily arXiv:2307.04729 [hep-th]}}.

\bibitem{Haring:2023zwu}
K.~H{\"a}ring and A.~Zhiboedov, ``{The stringy S-matrix bootstrap: maximal spin and superpolynomial softness},'' \href{http://dx.doi.org/10.1007/JHEP10(2024)075}{{\em JHEP} {\bfseries 10} (2024) 075}, \href{http://arxiv.org/abs/2311.13631}{{\ttfamily arXiv:2311.13631 [hep-th]}}.

\bibitem{McPeak:2023wmq}
B.~McPeak, M.~Venuti, and A.~Vichi, ``{Adding subtractions: comparing the impact of different Regge behaviors},'' \href{http://arxiv.org/abs/2310.06888}{{\ttfamily arXiv:2310.06888 [hep-th]}}.

\bibitem{Dong:2024omo}
Z.-Y. Dong, T.~Ma, A.~Pomarol, and F.~Sciotti, ``{Bootstrapping the chiral-gravitational anomaly},'' \href{http://dx.doi.org/10.1007/JHEP05(2025)114}{{\em JHEP} {\bfseries 05} (2025) 114}, \href{http://arxiv.org/abs/2411.14422}{{\ttfamily arXiv:2411.14422 [hep-th]}}.

\bibitem{Berman:2023jys}
J.~Berman, H.~Elvang, and A.~Herderschee, ``{Flattening of the EFT-hedron: supersymmetric positivity bounds and the search for string theory},'' \href{http://dx.doi.org/10.1007/JHEP03(2024)021}{{\em JHEP} {\bfseries 03} (2024) 021}, \href{http://arxiv.org/abs/2310.10729}{{\ttfamily arXiv:2310.10729 [hep-th]}}.

\bibitem{Berman:2024eid}
J.~Berman, H.~Elvang, N.~Geiser, and L.~L. Lin, ``{Bootstrapping Extremal Scalar Amplitudes With and Without Supersymmetry},'' \href{http://arxiv.org/abs/2412.13368}{{\ttfamily arXiv:2412.13368 [hep-th]}}.

\bibitem{Berman:2024wyt}
J.~Berman and H.~Elvang, ``{Corners and islands in the S-matrix bootstrap of the open superstring},'' \href{http://dx.doi.org/10.1007/JHEP09(2024)076}{{\em JHEP} {\bfseries 09} (2024) 076}, \href{http://arxiv.org/abs/2406.03543}{{\ttfamily arXiv:2406.03543 [hep-th]}}.

\bibitem{Berman:2025owb}
J.~Berman, H.~Elvang, and C.~Figueiredo, ``{Splitting Regions and Shrinking Islands from Higher Point Constraints},'' \href{http://arxiv.org/abs/2506.22538}{{\ttfamily arXiv:2506.22538 [hep-th]}}.

\bibitem{laurentgengoux2024invitationsingularfoliations}
C.~Laurent-Gengoux, R.~Louis, and L.~Ryvkin, ``An invitation to singular foliations,'' 2024.
\newblock \url{https://arxiv.org/abs/2407.14932}.

\bibitem{Mandelstam:1958xc}
S.~Mandelstam, ``{Determination of the pion - nucleon scattering amplitude from dispersion relations and unitarity. General theory},'' \href{http://dx.doi.org/10.1103/PhysRev.112.1344}{{\em Phys. Rev.} {\bfseries 112} (1958) 1344--1360}.

\bibitem{Correia:2020xtr}
M.~Correia, A.~Sever, and A.~Zhiboedov, ``{An analytical toolkit for the S-matrix bootstrap},'' \href{http://dx.doi.org/10.1007/JHEP03(2021)013}{{\em JHEP} {\bfseries 03} (2021) 013}, \href{http://arxiv.org/abs/2006.08221}{{\ttfamily arXiv:2006.08221 [hep-th]}}.

\bibitem{Correia:2021etg}
M.~Correia, A.~Sever, and A.~Zhiboedov, ``{Probing multi-particle unitarity with the Landau equations},'' \href{http://dx.doi.org/10.21468/SciPostPhys.13.3.062}{{\em SciPost Phys.} {\bfseries 13} no.~3, (2022) 062}, \href{http://arxiv.org/abs/2111.12100}{{\ttfamily arXiv:2111.12100 [hep-th]}}.

\bibitem{Paulos:2016fap}
M.~F. Paulos, J.~Penedones, J.~Toledo, B.~C. van Rees, and P.~Vieira, ``{The S-matrix bootstrap. Part I: QFT in AdS},'' \href{http://dx.doi.org/10.1007/JHEP11(2017)133}{{\em JHEP} {\bfseries 11} (2017) 133}, \href{http://arxiv.org/abs/1607.06109}{{\ttfamily arXiv:1607.06109 [hep-th]}}.

\bibitem{vanRees:2022zmr}
B.~C. van Rees and X.~Zhao, ``{Quantum Field Theory in AdS Space instead of Lehmann-Symanzik-Zimmerman Axioms},'' \href{http://dx.doi.org/10.1103/PhysRevLett.130.191601}{{\em Phys. Rev. Lett.} {\bfseries 130} no.~19, (2023) 191601}, \href{http://arxiv.org/abs/2210.15683}{{\ttfamily arXiv:2210.15683 [hep-th]}}.

\bibitem{vanRees:2023fcf}
B.~C. van Rees and X.~Zhao, ``{Flat-space Partial Waves From Conformal OPE Densities},'' \href{http://arxiv.org/abs/2312.02273}{{\ttfamily arXiv:2312.02273 [hep-th]}}.

\bibitem{Martin:1969ina}
A.~Martin, \href{http://dx.doi.org/10.1007/BFb0101043}{{\em {Scattering Theory: Unitarity, Analyticity and Crossing}}}, vol.~3.
\newblock Springer Berlin Heidelberg, Berlin, Heidelberg, 1969.
\newblock \url{https://doi.org/10.1007/BFb0101044}.

\bibitem{Eden:1966dnq}
R.~J. Eden, P.~V. Landshoff, D.~I. Olive, and J.~C. Polkinghorne, {\em {The analytic S-matrix}}.
\newblock Cambridge Univ. Press, Cambridge, 1966.

\bibitem{Math}
H.~Rosengren, ``{String theory amplitudes and partial fractions},'' \href{http://dx.doi.org/10.1007/s11139-025-01080-z}{{\em Ramanujan J.} {\bfseries 67} no.~2, (2025) 26}, \href{http://arxiv.org/abs/2409.06658}{{\ttfamily arXiv:2409.06658 [math.CA]}}.

\bibitem{Green:2019tpt}
M.~B. Green and C.~Wen, ``{Superstring amplitudes, unitarily, and Hankel determinants of multiple zeta values},'' \href{http://dx.doi.org/10.1007/JHEP11(2019)079}{{\em JHEP} {\bfseries 11} (2019) 079}, \href{http://arxiv.org/abs/1908.08426}{{\ttfamily arXiv:1908.08426 [hep-th]}}.

\bibitem{Chowdhury:2022obp}
D.~Chowdhury, P.~Haldar, and A.~Zahed, ``{Locality and analyticity of the crossing symmetric dispersion relation},'' \href{http://dx.doi.org/10.1007/JHEP10(2022)180}{{\em JHEP} {\bfseries 10} (2022) 180}, \href{http://arxiv.org/abs/2205.13762}{{\ttfamily arXiv:2205.13762 [hep-th]}}.

\end{thebibliography}\endgroup



\small
 
\end{document}